# Room-temperature oxygen transport in nano-thin Bi$_x$O$_y$Se$_z$ enables precision modulation of 2D materials

Zachariah Hennighausen[1,*], Bethany M. Hudak[2], Madeleine Phillips[2], Jisoo Moon[1], Kathleen M. McCreary[2], Hsun-Jen Chuang[3], Matthew R. Rosenberger[4], Berend T. Jonker[2], Connie H. Li,[2] Rhonda M. Stroud[2], and Olaf M. van 't Erve[2,*]

[1] NRC Postdoc Residing at the Materials Science and Technology Division, United States Naval Research Laboratory, Washington, D.C. 20375, USA
[2] Materials Science and Technology Division, United States Naval Research Laboratory, Washington, D.C. 20375, USA
[3] Nova Research, Inc., Alexandria, VA 22308, USA
[4] University of Notre Dame, Notre Dame, IN 46556, USA

**Abstract**

**Oxygen conductors and transporters are important to several consequential renewable energy technologies, including fuel cells and syngas production. Separately, monolayer transition metal dichalcogenides (TMDs) have demonstrated significant promise for a range of applications, including quantum computing, advanced sensors, valleytronics, and next-gen optoelectronics. Here, we synthesize a few nanometer-thick Bi$_x$O$_y$Se$_z$ compound that strongly resembles a rare $R\bar{3}m$ bismuth oxide (Bi$_2$O$_3$) phase, and combine it with monolayer TMDs, which are highly sensitive to their environment. We use the resulting 2D heterostructure to study oxygen transport through Bi$_x$O$_y$Se$_z$ into the interlayer region, whereby the 2D material properties are modulated, finding extraordinarily fast diffusion at room temperature under laser exposure. The oxygen diffusion enables reversible and precise modification of the 2D material properties by controllably intercalating and deintercalating oxygen. Changes are spatially confined, enabling submicron features (e.g. pixels), and are long-term stable for more than 221 days. Our work suggests few nanometer-thick Bi$_x$O$_y$Se$_z$ is a promising unexplored room-temperature oxygen transporter. Additionally, our findings suggest the mechanism can be applied to other 2D materials as a generalized method to manipulate their properties with high precision and submicron spatial resolution.**

\* Authors for correspondence, E-mail:  zachariah.hennighausen.ctr@mail.nrl.navy.mil; olaf.vanterve@nrl.navy.mil;

**Introduction**

Oxygen transporters and ion conductors – materials that facilitate rapid oxygen diffusion – enable key technologies that are expected to play a critical role as the world transitions to clean and renewable energy, including solid oxide fuel cells (SOFCs) and synthesis gas (syngas) production.[1–3] SOFCs are not subject to Carnot cycle limitations, efficiently producing electricity from direct fuel oxidation without moving parts. Additionally, they are scalable, modular, low-noise, and can be operated in reverse to produce syngas and oxygen. Oxygen transport membranes and oxygen carriers can also be applied in other processes to produce carbon-neutral syngas using biomass or solar heating.[4–6] Syngas, and its conversion to liquid hydrocarbons,[7] has demonstrated potential as a carbon-neutral fuel for long-duration and high-density energy storage, solving a host of challenges where current battery technology is impractical, including (1) intermittent and seasonal energy production; (2) long refueling times; and (3) heavy-duty and long-distance transportation (e.g., buses, trucks, trains, forklifts, and ships).[1]

Bismuth oxide-based materials are a leading candidate for advanced SOFCs[8–12] and oxygen separation membranes[13] and have demonstrated exceptional oxygen transport and ion conducting properties. Additionally, they have demonstrated promise for a spectrum of technologies, including photocatalysts,[14] multiferroics,[15] and gas sensors.[16] However, bismuth oxide-based materials are difficult to stabilize at sufficiently low sustainable operating temperatures.[8] Indeed, a critical limitation of SOFCs in general is the high operating temperature (~800°C), which decreases lifetime and increases start-up times, constraining greater adoption.[3]

In this work, we synthesize a novel few nanometer-thick oxygen transporter composed of a $Bi_xO_ySe_z$ nano-thin film that strongly resembles a rare $R\bar{3}m$ bismuth oxide ($Bi_2O_3$) phase, which, to the best of our knowledge, has never been studied in the nano-thin geometry. Our findings suggest that the $Bi_xO_ySe_z$ films facilitate rapid diffuse oxygen at room temperature under laser exposure. We mostly refer to the material under study in this work as $Bi_xO_ySe_z$, but STEM-EDS measurements suggest it contains so little selenium that it may be equally accurate to describe it as $R\bar{3}m$ $Bi_2O_3$. We synthesize this material by laser processing few-layer $Bi_2Se_3$ in an oxygen environment, which facilitates the steady incorporation of oxygen, removal of selenium, and a structural change. The precursor $Bi_2Se_3$ is grown using either chemical vapor deposition (CVD) or molecular-beam epitaxy (MBE), suggesting the process is economically scalable, providing a route to application.

We study oxygen transport through few nanometer-thick $Bi_xO_ySe_z$ and explore further applications by combining the material with monolayer transition metal dichalcogenides (TMDs). Like many 2D materials—which have made a notable impact across numerous disparate fields, including superconductivity,[17] quantum information science,[18] DNA sequencing,[19] catalysts,[20] transistors,[21] renewable energy,[22] and COVID-19 sensing[23]—monolayer TMDs are quite sensitive to their environment. They are particularly notable for their strong light-matter interaction, bright photoluminescence (PL), and tightly-bound excitons. In the heterostructures we construct, the oxygen transport through the $Bi_xO_ySe_z$ layer allows us to controllably shuttle oxygen into and out of the interlayer region, thereby reversibly modulating the 2D material's properties with high precision. The modulation is confined to the laser spot. As oxygen is intercalated (deintercalated), the PL is brightened (darkened), due to changing exciton recombination pathways. The PL evolution follows Fick's 2$^{nd}$ Law of Diffusion, enabling us to measure the oxygen diffusion speed through the $Bi_xO_ySe_z$. Our findings indicate oxygen diffuses very fast through $Bi_xO_ySe_z$ at room temperature under laser exposure (2.61e-18 $m^2s^{-1}$),[24,25] raising the prospect that it could advance low-temperature SOFC technology and improve syngas production. As an additional application, our work suggests $Bi_xO_ySe_z$ oxygen transport facilitates the precise manipulation of 2D materials properties on the submicron scale, raising the possibility for a host of spatially-selective and tunable properties, including long-lived interlayer excitons,[26] magnetism,[27] and ferroelectricity.[28]

**Experiment**

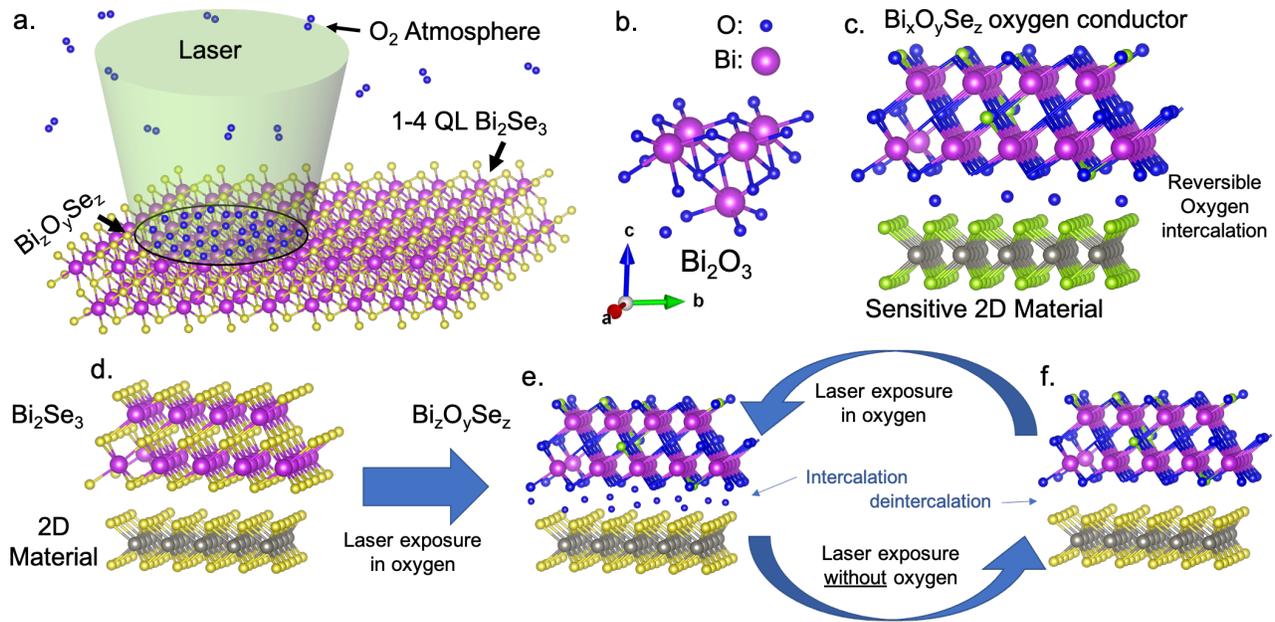

**Figure 1: Laser-oxygen exposure facilitates Bi$_x$O$_y$Se$_z$ transformation, enabling oxygen intercalation and deintercalation at 2D material interface.** (a) Schematic showing spatially selective transformation of few-quintuple layer (QL) Bi$_2$Se$_3$ into few-nanometer thin Bi$_x$O$_y$Se$_z$ using a laser in air at room temperature. During laser exposure, oxygen is incorporated and selenium is removed. (b) Regions of the laser exposed area are converted into a material that strongly resembles a rare $R\bar{3}m$ Bi$_2$O$_3$ phase. Bismuth oxides are a class of materials well known for their high oxygen ion conductance. (c) Bi$_x$O$_y$Se$_z$ can be used to precisely modulate the properties of 2D materials with high spatial selectivity. Regulating the quantity of intercalated oxygen between Bi$_x$O$_y$Se$_z$ and the 2D material modulates the coupling between the materials and the 2D material properties. (d-f) Summary of 2D heterostructure processing, and controlled intercalation-deintercalation cycling. While (d) Bi$_2$Se$_3$ is converted to (e) Bi$_x$O$_y$Se$_z$, oxygen simultaneously intercalates through Bi$_x$O$_y$Se$_z$, thereby decoupling the materials. (e-f) The oxygen can be controllably intercalated and deintercalated, thereby modulating the coupling between the materials.

Our findings suggest laser exposure in an oxygen atmosphere converts few-layer Bi$_2$Se$_3$ into a few-nanometer thin Bi$_x$O$_y$Se$_z$ compound that strongly resembles $R\bar{3}m$ Bi$_2$O$_3$ phase (Figure 1a). Figure 1b shows the $R\bar{3}m$ Bi$_2$O$_3$ phase. Previous work reported bulk rhombohedral Bi$_2$O$_3$, although it needed to be stabilized using dopants or a substrate.[9–11] Our work suggests $R\bar{3}m$ Bi$_2$O$_3$ is stable in ambient as a nano-thin film. Figure 1c-f illustrates Bi$_x$O$_y$Se$_z$ facilitating the transport of oxygen into the interlayer region, thereby modifying the 2D material environment and its coupling to Bi$_x$O$_y$Se$_z$. More specifically, when oxygen is absent (present) in the interlayer region, the materials are coupled (uncoupled), modifying the material's properties and exciton recombination pathways. Previous work showed that the interlayer coupling strength of layered materials can be modulated by intercalating oxygen or other compounds.[29–31] The sensitivity of the 2D material to the presence of intercalated oxygen allows us to quantify the oxygen transport through the Bi$_x$Se$_y$O$_z$ nano-thin film.

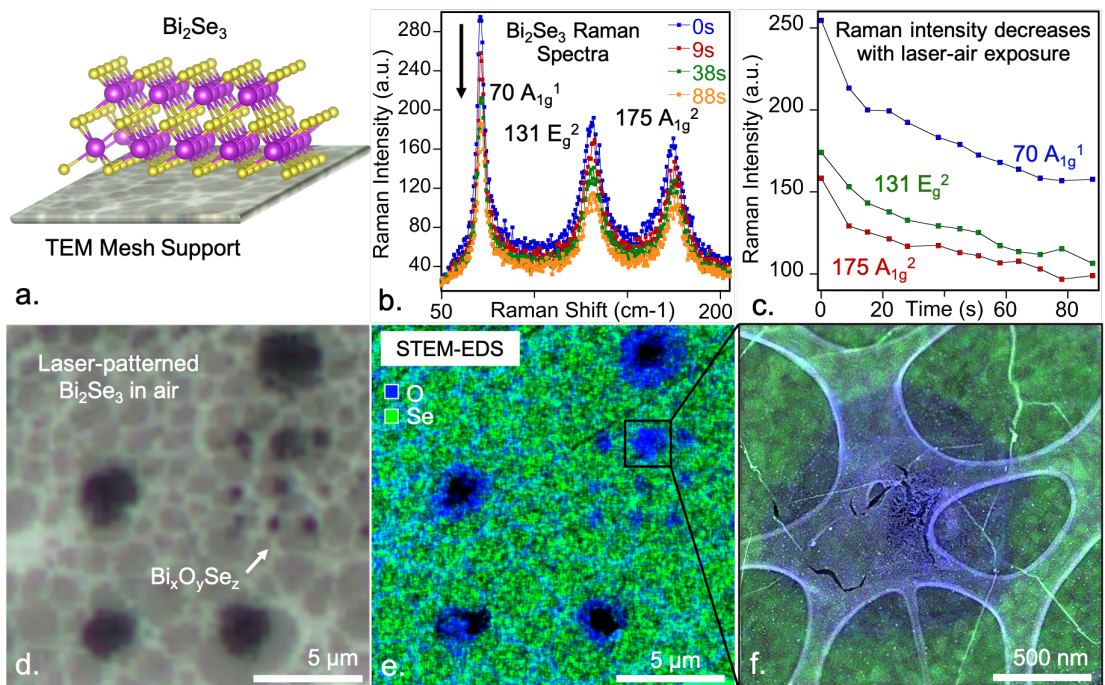

**Figure 2: Site-selective Bi$_x$O$_y$Se$_z$ fabrication on TEM mesh support reveals stoichiometric changes.** (a) Schematic showing monolayer Bi$_2$Se$_3$ on a TEM mesh support. (b)-(c) Raman spectra measured during laser exposure in air reveal the decrease of characteristic Bi$_2$Se$_3$ peaks, suggesting the material is being disrupted. (d) Optical image of Bi$_2$Se$_3$ on a TEM mesh support laser-patterned in air. Laser-oxygen exposure induces a prominent color change (grid of small dark spots). Large dark spots are laser-drilled holes created for ease of sample navigation. (e)-(f) Corresponding STEM-EDS maps. Where Bi$_2$Se$_3$ was exposed to the laser in air, the concentration of oxygen increases, while selenium decreases.

Bi$_x$O$_y$Se$_z$ was fabricated on a Cu mesh TEM grid with a lacey carbon support, and studied using Raman spectroscopy and scanning transmission electron microscopy energy-dispersive X-ray spectroscopy (STEM-EDS). Few-layer Bi$_2$Se$_3$ – the Bi$_x$O$_y$Se$_z$ precursor – was grown using molecular beam epitaxy (MBE) and transferred onto a TEM grid (see Figure 2a and methods). Raman spectroscopy reveals peaks characteristic of few-layer Bi$_2$Se$_3$ (Figure 2b). During laser-air exposure, all the peaks decreased in intensity (Figure 2b-c), suggesting the Bi$_2$Se$_3$ was being disrupted. To the best of our knowledge, the $R\bar{3}m$ Bi$_2$O$_3$ Raman modes have not been previously identified, where further work is required to definitively label possible Bi$_2$O$_3$ Raman modes. Figure 2d shows an optical image of Bi$_x$O$_y$Se$_z$ laser-patterned into Bi$_2$Se$_3$, where the white "webbing" is from the lacey carbon support of the TEM grid and the dark spots are Bi$_x$O$_y$Se$_z$. Using prolonged (>4 min) high-power laser (393 µW) and oxygen exposure, four holes were drilled into the Bi$_2$Se$_3$ to provide unambiguous identification of the region of interest in the STEM. A 3 x 3 grid was laser-patterned using lower powers (27.0 µW) and shorter durations (1-2 min). Our findings suggest Bi$_x$O$_y$Se$_z$ exhibits a distinct color from Bi$_2$Se$_3$, enabling laser-patterned areas to be easily identified under optical imaging.

Our experiment and theory findings suggest we are fabricating $Bi_xO_ySe_z$ at both the edges of the drilled holes and in the 3x3 grid of spots. Figure 2e shows the same region from Figure 2d with the oxygen (blue) and selenium (green) content mapped via STEM-EDS. At the edge of each of the holes, we see a clear increase in the oxygen content with a corresponding decrease in selenium. The laser-oxygen-exposed 3 x 3 grid is also readily identified by the increased oxygen content. Figure 2f shows a higher magnification STEM-EDS map of a single spot from the grid: the approximately 1 μm diameter circle contains a large amount of oxygen and small amount of selenium. The shape and size of this spot corresponds well with the laser spot size. Using the STEM-EDS map in Figure 2f, we calculate an 87% decrease in Se – from 23 at. % to 3 at. % – and a 47% increase in O – from 57 at. % to 84 at. % – between the surrounding film and the laser-exposed spot. See Section S1 for additional information about the STEM-EDS measurements.

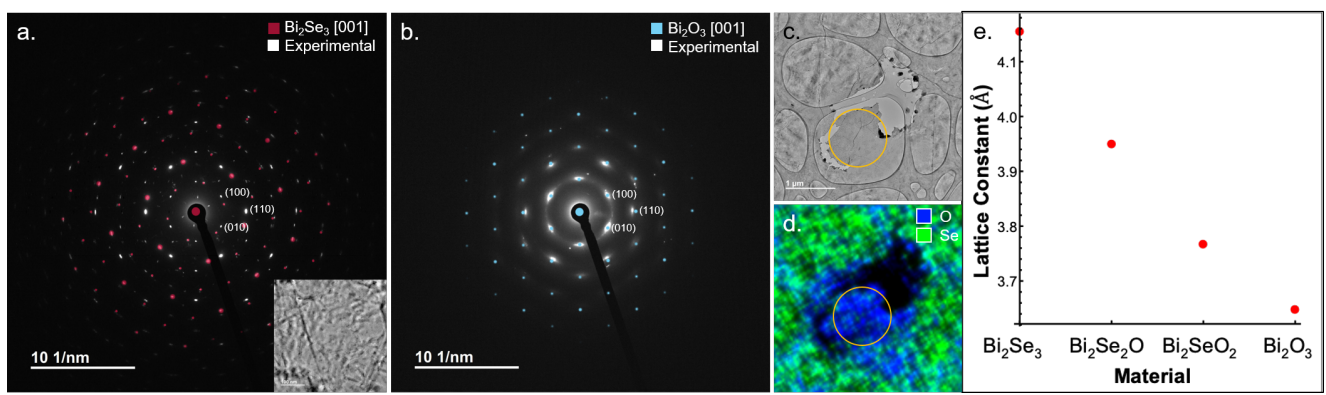

**Figure 3. Selected area electron diffraction (SAED) analysis of (a) pristine $Bi_2Se_3$ and (b) $Bi_xO_ySe_z$ formed through laser exposure. (**a) SAED from pristine multi-layer $Bi_2Se_3$, which matches well with the calculated $Bi_2Se_3$ pattern. Inset: HAADF image of pristine $Bi_2Se_3$. (b) SAED of $Bi_2Se_3$ region exposed to laser compared to calculated $Bi_2O_3$ (blue) diffraction patterns. The experimental pattern matches closely with the first-principles DFT calculated $Bi_2O_3$ structure. (c) HAADF image from which the SAED in (b) was acquired. Yellow circle indicates the position of the selected-area aperture. (d) EDS map of O (blue) and Se (green) from the region in (c). EDS indicates a decrease in Se and increase in O. (e) Lattice constants optimized using first-principles DFT of $Bi_2Se_3$ and related materials, as Se is systematically replaced with O, and $R\bar{3}m$ structure is maintained.

The change in O and Se content correlates with a change in the crystal lattice. Figure 3 shows selected area electron diffraction (SAED) taken from two regions of the film: a pristine $Bi_2Se_3$ region (Figure 3a) and a laser-oxygen exposed region (Figure 3b). The pristine $Bi_2Se_3$ region was located on the TEM grid several grid-squares away from the laser-exposed region and therefore we can confidently assume it was not impacted by the laser. The SAED pattern of the $Bi_2Se_3$ shows two sets of spots, indicating a twist or fold among the layers. As shown in Figure 3a, the SAED pattern matches well with the one quintuple layer $R\bar{3}m$ $Bi_2Se_3$ pattern calculated using first-principles density functional theory (DFT) (see methods). Figure 3b shows an SAED pattern from a laser-oxygen exposed spot. The yellow circles in Figure 3c-d show the position of the SAED aperture, and a significant increase in O can be seen in the STEM-EDS map (Figure 3d). The SAED pattern of this region has a smaller lattice constant than $Bi_2Se_3$, and the crystal structure matches well with a single quintuple layer of $R\bar{3}m$ $Bi_2O_3$, as computed in DFT. The SAED data combined with the STEM-EDS data strongly indicates that the laser-oxygen exposure of the $Bi_2Se_3$ is removing Se atoms and replacing them with O atoms, resulting in the production of a $Bi_xO_ySe_z$ compound with a reduced lattice constant that differs by <1% of $R\bar{3}m$ $Bi_2O_3$ (Section S1).

To further explore this possibility, we used DFT to optimize the lattice constants of a series of monolayer $Bi_2Se_3$ crystals in which the selenium atoms are systematically replaced with oxygen atoms (Figure 3e). The crystal structure of $Bi_2Se_3$ ($R\bar{3}m$) is maintained for each calculation. The structure that most nearly matches the SAED results is the monolayer $Bi_2O_3$. Inspection of the phonon modes of $Bi_2O_3$ with the $R\bar{3}m$ crystal structure shows that this structure is not predicted to be stable at 0K (Section S2). However, previous work experimentally demonstrated that bulk rhombohedral $Bi_2O_3$ using dopants and substrates[9–11] at non-cryogenic temperatures, raising the possibility that a combination of structural, chemical, and environmental factors is stabilizing the rare $Bi_2O_3$ phase in our experiments. For example, several other $Bi_2O_3$ phases, such as δ-$Bi_2O_3$, only stabilize above certain temperatures.[8]

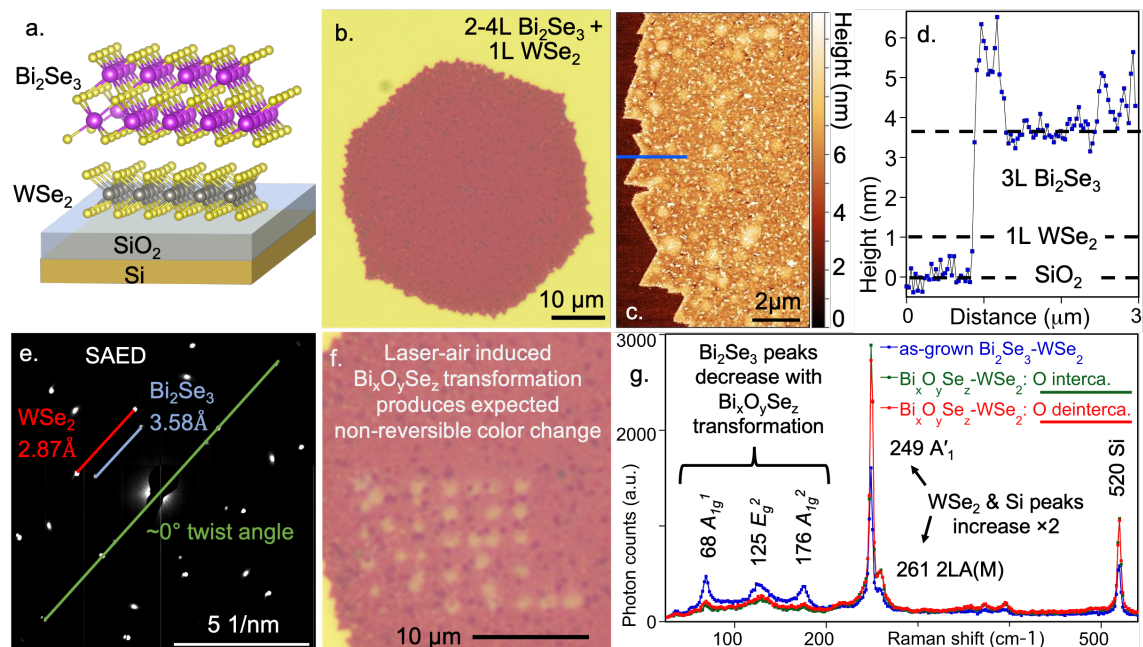

**Figure 4: Fabricating $Bi_xO_ySe_z$-$WSe_2$ 2D heterostructure.** (a) Schematic of a $Bi_2Se_3$-$WSe_2$ 2D heterostructure grown on $SiO_2$ using CVD. (b)-(e) As-grown $Bi_2Se_3$-$WSe_2$ 2D heterostructure. (b) Optical image. Yellow background is $SiO_2$. (c) AFM scan and (d) corresponding line profile showing 3-layers $Bi_2Se_3$ were grown on 1-layer $WSe_2$. (e) Selected area electron diffraction (SAED) image showing $Bi_2Se_3$ prefers to grow at a ~0° twist angle on $WSe_2$. (f) As-grown $Bi_2Se_3$-$WSe_2$ was laser-patterned in air with the letters "NRL", inducing spatially selective transformation into $Bi_xO_ySe_z$-$WSe_2$. $Bi_xO_ySe_z$ is confined to the laser spot, where the expected color change readily observed (see Figure 2d). The color change is due to $Bi_xO_ySe_z$ and independent of the concentration of intercalated oxygen. (g) Raman spectra of three configurations: as-grown $Bi_2Se_3$-$WSe_2$, and $Bi_xO_ySe_z$-$WSe_2$ with both oxygen intercalated and oxygen deintercalated. During $Bi_xO_ySe_z$ transformation, $Bi_2Se_3$ peaks significantly diminish (see Figure 2b-c). Conversely, the $WSe_2$ and Si peaks increase a factor ×2, possibly due to the greater optical transmission through $Bi_xO_ySe_z$.

We now turn to the study of our nano-thin $Bi_xO_ySe_z$ formed in heterostructures with monolayer TMDs. Numerous monolayer TMDs – including $WSe_2$ and $WS_2$ – have a direct band gap and tightly-bound excitons, which facilitate bright PL and strong light matter interaction. The monolayer nature enables the exciton's electric field lines to extend outside the material, making them sensitive to changes in the surrounding dielectric environment. They are promising materials for a spectrum of optoelectronic and advanced computing applications, including valleytronics,[32] twistronics,[33] spintronics,[34] and quantum information sciences.[18] We take advantage of the sensitivity of monolayer TMD excitons to study the oxygen transport through our $Bi_xO_ySe_z$ material and to propose an application for our material as an aid in modulating 2D materials properties.

Figure 4 demonstrates spatially-selective $Bi_xO_ySe_z$-$WSe_2$ 2D heterostructure fabrication through laser exposure in air of a $Bi_2Se_3$-$WSe_2$ 2D heterostructure. $Bi_2Se_3$-$WSe_2$ is grown using CVD, where monolayer $WSe_2$ is first grown on $SiO_2$/Si, followed by $Bi_2Se_3$ on top (see methods). Figure 4a-b is a schematic and optical image, respectively, of as-grown $Bi_2Se_3$-$WSe_2$ on an $SiO_2$/Si substrate. Figure 4c-d shows an atomic force microscope (AFM) scan and corresponding line profile, respectively, showing three layers of $Bi_2Se_3$ were grown on monolayer $WSe_2$. Figure 4e shows an SAED image of $Bi_2Se_3$-$WSe_2$, where the well-formed spots suggest long-range crystallinity. Lattice constants were measured as 4.13 Å and 3.31 Å, respectively. The $Bi_2Se_3$ grows at a near ~0° twist angle, suggesting material interaction is sufficiently strong to steer growth dynamics. Figure 4f shows $Bi_xO_ySe_z$-$WSe_2$ patterned into $Bi_2Se_3$-$WSe_2$ as the letters "NRL", where the expected color change due to $Bi_xO_ySe_z$ is observed (previously shown in Figure 2d). Figure 4g shows the Raman spectrum of as-grown $Bi_2Se_3$-$WSe_2$, as well as $Bi_xO_ySe_z$-$WSe_2$ with both oxygen intercalated and oxygen deintercalated. In agreement with Figure 2b-c, $Bi_2Se_3$ peaks decrease, suggesting the crystal is being altered during the first laser-oxygen exposure. Conversely, $WSe_2$ and Si peaks increase, suggesting the optical transmission through $Bi_xO_ySe_x$ increases. Although the PL intensity and peak position undergo dramatic shifts between the oxygen intercalated and deintercalated states (discussed later), no detectable change in Raman spectrum is observed, suggesting PL modification is due to electronic changes at the interface that only negligibly affect the phonon modes.

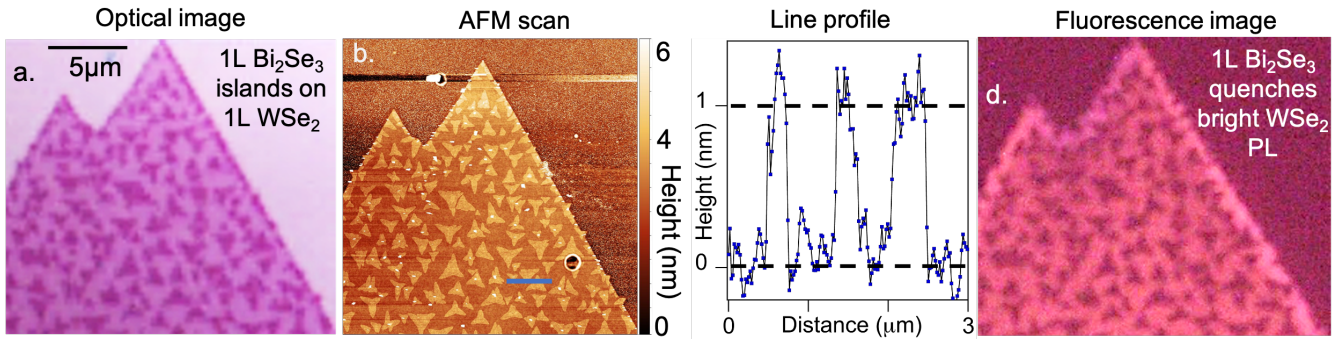

**Figure 5: Monolayer Bi$_2$Se$_3$ quenches bright monolayer WSe$_2$ PL.** (a) Optical image showing triangular Bi$_2$Se$_3$ crystals grown on monolayer WSe$_2$. Bi$_2$Se$_3$ triangles grow aligned with WSe$_2$, in agreement with SAED. (b) AFM scan and (c) corresponding line profile shows Bi$_2$Se$_3$ triangles are one quintuple layer (1nm) tall. Blue line corresponds to the location of the line profile. (d) Fluorescence image showing monolayer Bi$_2$Se$_3$ quenches the bright WSe$_2$ PL, suggesting the strong electronic coupling induces a non-radiative exciton recombination pathway.

Figure 5 shows growing one Bi$_2$Se$_3$ layer on monolayer WSe$_2$ quenches the bright PL, suggesting the clean interface and a strong electronic coupling facilitates non-radiative exciton recombination. This is in agreement with theory calculations that predict hybridized states at the Gamma point facilitate a non-radiative decay pathway for carriers excited at K in WSe$_2$ (See Section S3). Figure 5a is an optical image of monolayer Bi$_2$Se$_3$ triangular crystals (dark purple) grown on monolayer WSe$_2$ (light purple) on an SiO$_2$ substrate (white-pink). The well-formed triangular shapes suggest both materials are crystalline, in agreement with SAED measurements (Figure 4e). Figure 5b-c are an AFM scan and corresponding line profile, respectively, showing the Bi$_2$Se$_3$ grew as a monolayer (a single quintuple layer). Figure 5d is a fluorescence image showing the PL intensity contrast between monolayer WSe$_2$ and a Bi$_2$Se$_3$-WSe$_2$ 2D heterostructure. While monolayer WSe$_2$ exhibits the expected bright PL, a single layer of Bi$_2$Se$_3$ quenches the PL nearly completely, behavior that has been observed in other Bi$_2$Se$_3$-TMD 2D heterostructures.[35–38] Previous work suggests the strong interlayer coupling in Bi$_2$Se$_3$-TMD 2D heterostructures induces the formation of a pure electronic lattice at the interface, encouraging hybridization and non-radiative excitonic transitions.[39]

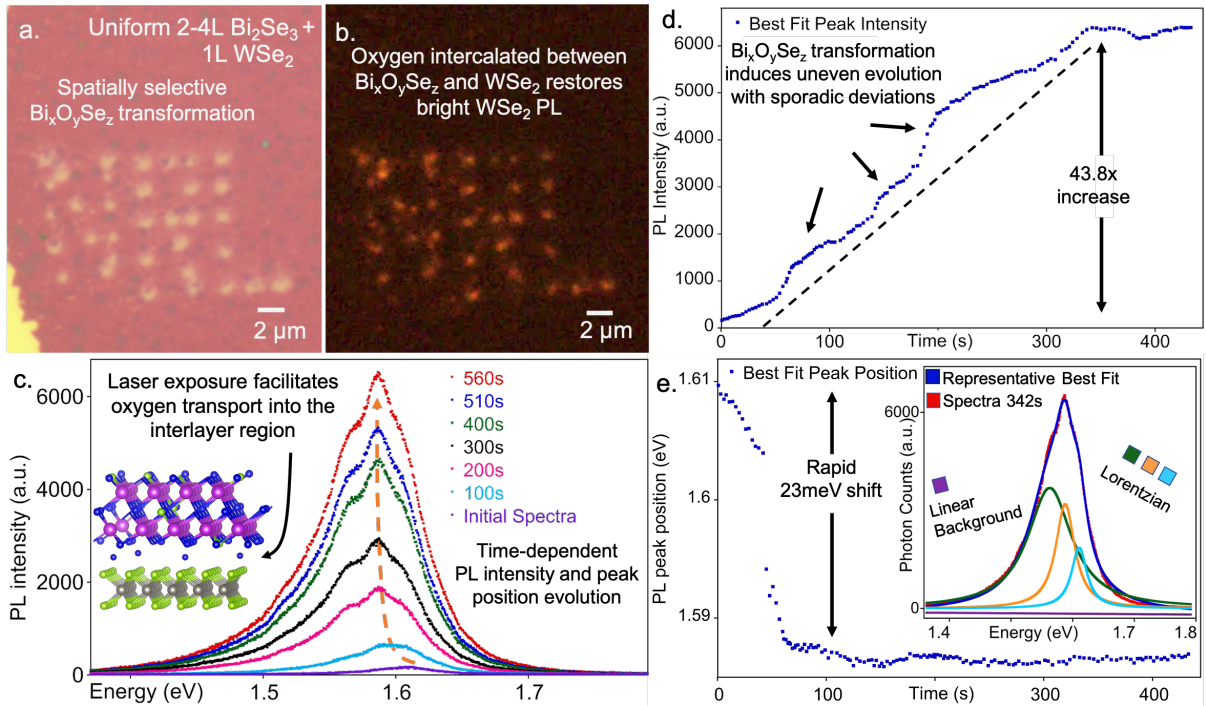

**Figure 6. Bi$_x$O$_y$Se$_z$ transformation facilitates oxygen intercalation and restoration of WSe$_2$ PL.** (a) Optical and (b) fluorescence images of Bi$_2$Se$_3$-WSe$_2$ 2D heterostructure laser-patterned in air with the letters "NRL", demonstrating localized Bi$_x$O$_y$Se$_z$ transformation with submicron resolution (740nm). (a) shows the expected color change, while (b) shows the WSe$_2$ PL is recovered where Bi$_x$O$_y$Se$_z$ is present. (c) PL spectra at multiple exposure times. The PL spectra steadily increases in intensity, changes shape, and shifts peak position with increasing laser exposure in air. As Bi$_2$Se$_3$ is transformed into Bi$_x$O$_y$Se$_z$, oxygen is simultaneously transported into the interlayer region, which decouples the materials, enabling WSe$_2$ independence and PL restoration (inset). (d)-(e) Evolution of (d) PL intensity and (e) peak position with increasing laser exposure. The PL intensity increases a factor of 43.8×, while the peak position redshifts by 23 meV. (d) The PL intensity line shape contains numerous sporadic deviations, due to the active transformation from Bi$_2$Se$_3$ to Bi$_x$O$_y$Se$_z$. (e) The PL peak position redshifts rapidly by 23 meV after ~60 s of exposure, and then remains mostly constant. This rapid shift is reversible through oxygen deintercalation (discussed later). 143 spectra were collected over 7.2 minutes, enabling a high-resolution understanding of system evolution. Each data point is extracted from robust fitting to a spectrum using three Lorentzian functions and a linear background (see inset for representative fitting and Section S4 for additional information).

Figure 6 demonstrates that the initial laser-oxygen exposure that induces transformation into Bi$_x$O$_y$Se$_z$ transformation also restores the WSe$_2$ PL. We believe that two things occur during this first laser-oxygen exposure: (1) Bi$_2$Se$_3$ is transformed into Bi$_x$O$_y$Se$_z$, and (2) oxygen is transported into the interlayer region, which decouples the materials, enabling radiative exciton recombination. The conduction band minimum of $R\bar{3}m$ Bi$_2$O$_3$ is very similar in shape and orbital contribution to that of Bi$_2$Se$_3$ (Section S5), so in the absence of oxygen intercalation at the interface, we would expect hybridization between the Bi$_x$O$_y$Se$_z$ and WSe$_2$ to form a non-radiative decay pathway, in the same way as in the Bi$_2$Se$_2$-WSe$_2$ heterostructure (Section S3).

Figure 6a-b show optical and fluorescence images, respectively, of Bi$_x$O$_y$Se$_z$ patterned as the letters "NRL" into a Bi$_2$Se$_3$-WSe$_2$ 2D heterostructure using laser exposure in air. This produces a purple-to-white

color change due to the structural transformation, and site-selective WSe$_2$ PL recovery with submicron (740 nm) resolution due to local oxygen intercalation. Figure 6c shows PL spectra at multiple laser-air exposure times, revealing time-dependent evolution of the PL intensity, peak position, and spectrum shape. Each spectrum contains a peak along with two shoulders, suggesting multiple radiative recombination pathways are present, possibly due to various exciton species[40,41] and phonon side bands[42] (discussed further in Section S4). Figure 6d-e show the PL intensity and peak position evolution, respectively, where each data point is extracted using a robust fitting algorithm written in Python Spyder software, enabling low-error measurements.[43] Each spectrum is fit with three Lorentzian functions and a linear background, where visual observation and low optimization uncertainty suggest a good fit (see inset for representative fitting, and Section S4 for additional information). The PL intensity evolution is continuous, steadily increasing 43.8× over 342 s (118 data points), demonstrating a continuum of PL values can be accessed. The PL peak position rapidly redshifts by 23 meV and changes shape during the first 65 s (23 data points), before plateauing and maintaining shape for the remainder of the exposure, a marked departure from the PL intensity evolution.

We believe the sporadic deviations (i.e., bumps or indentations that depart from an underlying order) in the PL intensity are likely due to chemical and structural changes during the Bi$_2$Se$_3$ to Bi$_x$O$_y$Se$_z$ transformation, where laser spot spatial inhomogeneity might be a contributing factor. These deviations are not observed in follow-on oxygen deintercalation-intercalation cycles (discussed later). More specifically, in follow-on exposures, the PL intensity reliably modulates higher and lower, following Fick's 2$^{nd}$ Law of Diffusion closely (discussed later). If the increase in PL intensity is indeed due to oxygen intercalation between layers, this difference in line shape suggests that oxygen transport to the interlayer region is more efficient through the Bi$_x$O$_y$Se$_z$ material than through the Bi$_2$Se$_3$ or partially structurally-transformed material.

As the PL is recovered during the initial laser-air exposure, the spectrum shape appears to evolve from a single exciton fit into a multi exciton fit. The relative intensity, peak position, and FWHM for each Lorentzian evolve as well, thereby changing the spectrum shape (Figure S6). Jumps in peak position also occur in subsequent oxygen intercalation and deintercalation cycles, after structural transformation is complete (Figure 7b). Together, our findings suggest the exciton recombination pathways are modified as oxygen is intercalated. Previous work found monolayer WSe$_2$ to have numerous exciton recombination pathways, as well as phonon side bands, which affect spectra shape.[41,44] Supporting information contains two movies that use all the raw data to display the PL evolution. Control experiments of bare monolayer WSe$_2$ (without Bi$_x$O$_y$Se$_z$) under laser-air exposure showed only comparatively very minor changes over 6 min of exposure (<6% increase), demonstrating that Bi$_x$O$_y$Se$_z$ is required to modulate the TMD PL (Section S6).

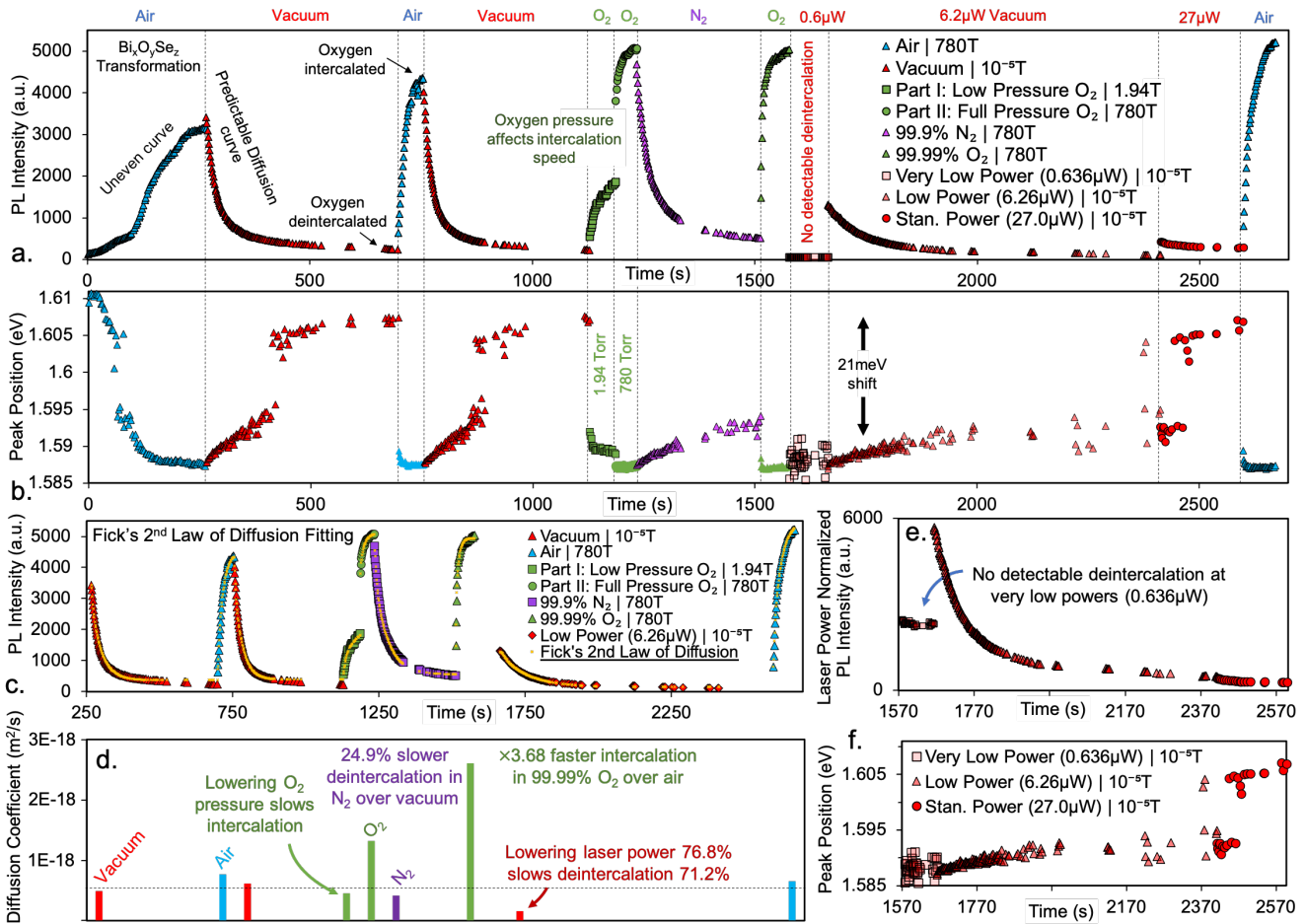

**Figure 7: Controlled oxygen intercalation and deintercalation enables reversible precision manipulation of 2D materials, and measurement of oxygen diffusion through Bi$_x$O$_y$Se$_z$.** (a) PL intensity and (b) peak position evolution from a Bi$_x$O$_y$Se$_z$-WSe$_2$ 2D heterostructure subject to multiple atmospheres and laser powers, where the diffusion speed is dependent on oxygen partial pressure and laser power. (a) The first laser-air exposure shows an uneven PL intensity curve, due to the Bi$_2$Se$_3$ to Bi$_x$O$_y$Se$_z$ transformation. In contrast, subsequent laser-atmosphere exposures follow Fick's 2$^{nd}$ Law of Diffusion closely (see (c)). (b) Changes to the peak position can be reversed. (c) Data from (a) replotted and fit with Fick's 2$^{nd}$ Law of Diffusion curves, enabling diffusion coefficients to be extracted. (d) Corresponding diffusion coefficients plotted. At room-temperature under laser exposure the material demonstrates very fast diffusion.[24,25] (e)-(f) Laser-power data replotted from (a) and (b), respectively, showing very low powers (0.636µW) have no detectable effect, while low powers (6.26µW) produce much slower diffusion. More specifically, lowering laser power 76.8%, slows deintercalation 71.2%. (f) The PL peak position only reliably returns to the initial value after the standard powers are applied.

Figure 7 demonstrates that the PL intensity can be repeatably increased and decreased by intercalating and deintercalating oxygen, where diffusion speeds are atmosphere, pressure, and laser-power dependent. Figure 7a-b show the PL intensity and peak position evolution, respectively, from a $Bi_xO_ySe_z$-$WSe_2$ 2D heterostructure exposed to multiple atmospheres, pressures, and laser powers. The PL intensity and peak position evolution is dependent on the oxygen partial pressure and the laser power, which affect the direction and speed of diffusion. The PL intensity (peak position) is increased (red-shifted) in an atmosphere containing oxygen, but decreased (blue-shifted) when oxygen is absent. The first laser-air exposure shows an uneven PL intensity curve, due to the $Bi_2Se_3$ to $Bi_xO_ySe_z$ transformation. In contrast, PL measured during subsequent laser-atmosphere exposures follows Fick's 2nd Law of Diffusion closely (Figure 7c), without the sporadic deviations observed in the first laser-oxygen exposure (Figure 6d). Changes to the peak position are also reversible, where the prominent rapid shift in peak position is consistently observed at low PL intensities. We suspect that when sufficient oxygen is deintercalated, $Bi_xO_ySe_z$ begins to couple to $WSe_2$, thereby strengthening certain higher-energy exciton recombination pathways. We demonstrate that a continuum of PL values can be reliably accessed, suggesting the oxygen intercalation can be controlled with high precision. Fluorescence imaging suggests the changes are stable at room temperature in $N_2$ for 221 days (Section S7). Together, the technology demonstrates spatially-selective and tunable intensities with write-read-erase-reuse capability and long-term stability.

Fick's 2nd Law of Diffusion – an equation used to describe the change in concentration with respect to time and space – fits the PL intensity data well (Figure 7c),[45] supporting our argument that PL is restored due to oxygen transport into the interlayer region, and enabling diffusion coefficients to be extracted for the $Bi_xO_ySe_z$ material (Figure 7d). We make the ansatz that the PL intensity is proportional to the concentration of oxygen in the interlayer region, and assume a fixed concentration boundary condition at the atmosphere and an impermeable boundary condition at the TMD interface. More specifically, we assume the concentration of oxygen in the atmosphere is unaffected (i.e., fixed concentration), and that no oxygen passes into or through the TMD (i.e., impermeable boundary).

When applying the boundary conditions, Fick's 2nd Law of Diffusion reduces to Eq. (1), where $c$ is concentration, $L$ is the thickness of the material (i.e., $Bi_xO_ySe_z$ + interlayer region thickness), $D$ is the diffusion coefficient, $h$ is the mass transfer coefficient, $K$ is the distribution coefficient, $x$ is the distance from $WSe_2$, and $t$ is the time. We set L = 3.32 nm based on AFM measurements (Figure 4d) and theory calculations (Sections S3), which show 3nm of $Bi_2Se_3$ and predict a 0.32nm thick interlayer region, respectively. Since we make the ansatz that the PL intensity is proportional to the concentration of oxygen in the interlayer region, we set $x = 0$ nm. We fit Eq. (1) to the experimental data using an algorithm written in Python Spyder software, allowing the paramters $c_{air}$, $c_{interlayer}$, $D$, and the ratio $h/k$ to vary (see methods). The $h/k$ ratio accounts for finite oxygen mass transfer from the atmosphere into $Bi_xO_ySe_z$, as well as differences in oxygen concentration at steady state between the atmosphere and $Bi_xO_ySe_z$. The finte $h/k$ value suggests a gradient forms at the interface between the atmosphere and $Bi_xO_ySe_z$. An amplifying discussion of Fick's 2nd Law of Diffusion and a partial derivation using the chosen boundary conditions is shown in Section S7.

$$c(x,t) = c_{air} - (c_{air} - c_{Initial}) \sum_{n=1}^{\infty} C_n \exp(-\zeta_n^2 F) \cos\left(\zeta_n \frac{x}{L}\right) \quad (1)$$

$$F = \frac{tD}{L^2}; \quad C_n = \frac{4\sin(\zeta_n)}{2\zeta_n + \sin(2\zeta_n)}; \quad \zeta_n \tan(\zeta_n) = B; \quad B = \frac{hL}{KD}$$

Considering that the measurements were completed at room-temperature, the material demonstrates exceptionally fast diffusion (2.61e-18 m$^2$s$^{-1}$),[24,25] suggesting it is a promising low-temperature oxygen transporter and potentially an oxygen ion conductor.[46] A low-temperature laser was used, and no temperature-dependent WSe$_2$ response was detected, suggesting laser heating is minimal (<20 °C change).[47] For comparison, current SOFC technology operates in excess of 800 °C, which impedes wide-spread adoption.

Figure 7e-f show laser-power-dependent data replotted from Figure 7a-b, respectively. The heterostructure starts with oxygen intercalated and is laser-vacuum exposed, suggesting oxygen should gradually deintercalate. Very low powers (0.636 µW) have no detectable effect on the PL spectrum (discussed further in Section S9), while low powers (6.26 µW) produce much slower diffusion compared to standard powers (27.0 µW). More specifically, lowering laser power 76.8% slows deintercalation 71.2%. As the intensity plateaus toward low values, the PL peak position exhibits quasi-binary behavior, vacillating between the higher (1.605 eV) and lower (1.592 eV) energy states before remaining at the higher value. This behavior is also observed in the 1st and 2nd vacuum evolutions, suggesting the radiative recombination pathways and excitons undergo a notable change at the end of oxygen deintercalation.

We also fabricated $Bi_xO_ySe_z$ on top of monolayer WS$_2$, and demonstrated that the WS$_2$ PL intensity and peak position could be reversibly modulated in a manner comparable to $Bi_xO_ySe_z$-WSe$_2$ (Section S10), establishing that the effect is not confined to WSe$_2$. After the initial laser-oxygen exposure, PL intensity measured in all subsequent exposures followed Fick's 2nd Law of Diffusion closely, suggesting the controlled oxygen transport and intercalation under laser exposure is universal. Notably, the diffusion constants measured in the Bi$_x$O$_y$Se$_z$-WS$_2$ system are comparable to those measured in Bi$_x$O$_y$Se$_z$-WSe$_2$ (see Table S2), despite the fact that the growth recipes and underlying TMDs were different. This supports our claim that we are measuring very fast room temperature diffusion speeds through the Bi$_x$O$_y$Se$_z$ itself and not a heterostructure specific quantity. Previous works showed laser-air exposure of various Bi$_2$Se$_3$-TMD 2D heterostructures could increase PL intensity; however, the mechanism was not understood, and a chemical modification of Bi$_2$Se$_3$ was considered unlikely.[35–37] Together, the Bi$_x$O$_y$Se$_z$-WS$_2$ experiments substantially reinforce our claims that Bi$_x$O$_y$Se$_z$ is a promising room-temperature oxygen transporter, and that it can be harnessed to precisely transport oxygen, regardless of the substrate.

**Discussion**

Raman data (Figure 4g) and optical microscopy images (Fig 4f) suggest that the structural change from Bi$_2$Se$_3$ to Bi$_x$Se$_y$O$_z$ occurs during the first laser-air exposure of the material and remains unchanged thereafter. In contrast, the PL intensity cycles with the partial pressure of oxygen in the environment. This suggests that the changes in PL are dependent on an interaction with oxygen that does not structurally alter either layer in the heterostructure. This cycling behavior of the PL is consistent with oxygen intercalating and deintercalating from between the two layers, in agreement with previous work.[48] In pristine Bi$_2$Se$_3$-WSe$_2$, the WSe$_2$ PL is quenched (Figure 5d) due to hybridization between the WSe$_2$ valence bands and the Bi$_2$Se$_3$ conduction bands, leading to a non-radiative decay pathway for excited carriers in the WSe$_2$. When oxygen is intercalated between the layers, the layers are physically separated, and the two layers no longer hybridize, suggesting a radiative decay pathway is primary (Section S3). Supercells of zero-degree aligned Bi$_2$O$_3$-WSe$_2$ bilayers are prohibitively large for computing similar band hybridization in the structurally altered heterostructures. However, the band alignment of Bi$_2$O$_3$ and WSe$_2$ is qualitatively the same as the Bi$_2$Se$_3$-WSe$_2$ band alignment, and the orbital weight at the conduction band minimum in both Bi$_2$Se$_3$ and Bi$_2$O$_3$ is on the Bi p$_z$ orbitals (Section S5). Taken together, these facts suggest that the intercalation of oxygen between Bi$_x$O$_y$Se$_z$ and WSe$_2$ layers would lead to similar changes in hybridization—and thus similar changes in PL—as are seen in the pristine Bi$_2$Se$_3$-WSe$_2$ heterostructure.

The evolution of PL intensity in the Bi$_x$O$_y$Se$_z$-WSe$_2$ heterostructure follows Fick's 2$^{nd}$ Law of Diffusion, enabling us to measure the oxygen diffusion speed through the Bi$_x$O$_y$Se$_z$. Our findings indicate oxygen diffuses very fast through Bi$_x$O$_y$Se$_z$ at room temperature under laser exposure (2.61e-18 m$^2$s$^{-1}$),[24,25] raising the prospect that it is a low-temperature oxygen ion conductor. Although other materials and Bi$_2$O$_3$ phases (e.g., δ-Bi$_2$O$_3$) have demonstrated much faster diffusion, they generally require temperatures in excess of 500°C to be stabilized and efficiently operate, limiting material lifetime and widespread adoption. Although our measured Bi$_x$O$_y$Se$_z$ values (Figure 7d) are several orders of magnitude greater than other room-temperature oxygen transporters (~9.2e-22 m$^2$s$^{-2}$),[25] our experiment required a low-power laser to facilitate diffusion.

**Conclusion**

Materials that enable rapid oxygen diffusion are becoming increasingly important as the world transitions to renewable energy. Oxygen transport membranes and oxygen ion conductors enable several technologies that manufacture carbon-neutral syngas and efficiently convert it into energy, solving a host of challenges where current battery technology is impractical, including long-duration and high-density energy storage.[1–3] A low-temperature oxygen ion conductor breakthrough could resolve the critical limitation of excessive SOFC operating temperatures, facilitating greater adoption of the technology and renewable energy.[1–3] We synthesize a few nanometer-thick Bi$_x$O$_y$Se$_z$ compound that closely resembles a rare $R\bar{3}m$ bismuth oxide (Bi$_2$O$_3$) phase, which is able to rapidly transport oxygen at room-temperature under laser exposure (2.61e-18 m$^2$s$^{-1}$). To measure the oxygen transport speed, we combine Bi$_x$O$_y$Se$_z$ with monolayer TMDs, which have a strong light-matter coupling, bright PL, and are sensitive to the surrounding environment. As oxygen is transported through the Bi$_x$O$_y$Se$_z$, it fills the interlayer region, decoupling the TMD from the Bi$_x$O$_y$Se$_z$, and modulating the monolayer TMD properties and brightening the PL. The oxygen can be controllably intercalated (deintercalated) to brighten (darken) the PL intensity, where the diffusion profile follows Fick's 2$^{nd}$ Law of Diffusion. Using laser-patterning, the diffusion can be controlled with submicron spatial resolution, where changes are stable for more than 221 days. Our work suggests nano-thin Bi$_x$O$_y$Se$_z$ is a promising unexplored room-temperature oxygen transporter, raising the prospect that it could advance low-temperature SOFC technology and improve syngas production, thereby facilitating renewable energy. Additionally, Bi$_x$O$_y$Se$_z$ and the interlayer oxygen intercalation mechanism can be applied generally to 2D materials to precisely transport oxygen to the 2D material surface on the submicron scale, thereby manipulating the properties. Depending on the material's underlying characteristics, this raises the possibility for spatially-selective and tunable magnetism,[27] long-lived interlayer excitons,[26] ferroelectricity,[28] and integrated quantum photonics.[49] Finally, if the transport properties are modulated across large ranges comparable to the PL observed herein, laser-written p-n junctions, neuromorphic computing schemes,[50,51] and multistate optoelectronic devices could also be possible.

**Methods**
**Material Growth – Few-layer $Bi_2Se_3$:** The $Bi_2Se_3$ films were grown on 10 × 10 mm² c-plane (0001) sapphire ($Al_2O_3$) substrates using molecular beam epitaxy (MBE) with base pressure below $5 \times 10^{-10}$ Torr. The substrates were initially annealed ex-situ at 1,000 °C under the atmospheric pressure, and ozone cleaned in-situ under 200 Torr of oxygen pressure. It is then annealed at 600 °C for 20 min in the ultra-high vacuum MBE chamber. Individual sources of high-purity (99.999%) Bi and Se were evaporated from standard effusion cells during the film growth. Se flux was maintained at least ten times higher than Bi's to minimize Se vacancies. To obtain an atomically sharp interface between the $Bi_2Se_3$ layer and the substrate, we adopted the two-step growth scheme.[52] First, the initial 3 QL $Bi_2Se_3$ is grown at 170 °C. It is slowly annealed to 300 °C, and followed by deposition of the remaining 2QL $Bi_2Se_3$ layers. 5QL $Bi_2Se_3$ was grown. AFM scan of the as-grown $Bi_2Se_3$ is shown in Section S11.

**Material Growth – $Bi_2Se_3$-TMD 2D Heterostructure:** Monolayer TMDs are synthesized at ambient pressure in 2-inch diameter quartz tube furnaces on $SiO_2$/Si substrates (275 nm thickness of $SiO_2$). Separate dedicated furnaces are used for the growth of $WS_2$ and $WSe_2$ to prevent cross contamination. Prior to use, all $SiO_2$/Si substrates are cleaned in acetone, IPA, and Piranha etch ($H_2SO_4+H_2O_2$) then thoroughly rinsed in DI water. At the center of the furnace is positioned a quartz boat containing ~1g of $WO_3$ powder. Two $SiO_2$/Si wafers are positioned face-down, directly above the oxide precursor. A separate quartz boat containing the appropriate chalcogen powder (S or Se) is placed upstream, outside the furnace-heating zone, for the synthesis of $WS_2$ or $WSe_2$. The upstream $SiO_2$/Si wafer contains perylene-3,4,9,10-tetracarboxylic acid tetrapotassium salt (PTAS) seeding molecules, while the downstream substrate is untreated. The hexagonal PTAS molecules are carried downstream to the untreated substrate and promote lateral growth of the monolayer TMD. Pure argon (65 sccm) is used as the furnace heats to the target temperature. Upon reaching the target temperature in the range of 825 to 875 °C, 10 sccm $H_2$ is added to the Ar flow and maintained throughout the 10-minute soak and subsequent cooling to room temperature.

$Bi_2Se_3$ was grown on top of the monolayer TMDs (i.e., $WS_2$ and $WSe_2$) using chemical vapor deposition (CVD) in a two-zone furnace with a 2" quartz tube. High-purity $Bi_2Se_3$ flakes are ground using a mortar and pestle into a fine dust. The powdered $Bi_2Se_3$ is placed in a ceramic boat and inserted into the furnace's quartz tube, and pushed into the center of the furnace's first zone. The monolayer TMD, which is on an $SiO_2$ substrate, is placed downstream of the $Bi_2Se_3$ into the center of the furnace's second zone. The furnace is pumped down to ~20mTorr. An argon (Ar) carrier gas is flown into the furnace at 80sccm. The $Bi_2Se_3$ is heated to 520°C (550°C), and the WS$_2$ (WSe$_2$) are heated to 210°C (245°C). The ramp rate is ~55°C/min, and the total growth is 27 min (14 min).

**Transmission Electron Microscopy:** Few-layer Bi$_2$Se$_3$ was transferred onto a Cu mesh TEM grid with lacey carbon support using wet transfer,[53] where the Bi$_2$Se$_3$ is released from the substrate using buffered hydrofluoric acid (HF), rinsed in five water baths, and lifted onto the TEM grid. STEM-EDS was performed in a Nion UltraSTEM-X equipped with a Bruker XFlash EDS detector operated at 60 kV with a nominal beam current of 30 pA. Bruker Esprit 2.0 software was used for STEM-EDS analysis and quantification. SAED was performed using a JEOL 2200FS operated at 200 kV with a 1 μm selected-area aperture. CrystalMaker and SingleCrystal software were used for fitting the SAED patterns.

**Atmospheric measurements:** Atmospheric measurements were conducted inside a Janis ST-500 with vacuum, air, N$_2$, and O$_2$ connections. Unless stated otherwise, we used a 27.0μW 532nm laser without interruption at a 50x long working distance objective (1.15μW/μm$^2$).

**Photoluminescence Spectra computational analysis and fitting:** All code as written in Python using the Spyder integrated development environment (IDE). Spyder belongs to the MIT License and is distributed through the Anaconda environment. Each spectrum is time stamped, and our algorithm extracted this info. The curve_fit() function with a variety of initial values and boundary conditions were used to verify fit robustness. The fitting was corroborated using cross-validation, where we uniformly removed 20% of the data points from a spectrum and repeated fitting. No notable changes to the fitting were detected, suggesting noise is not skewing the fit.

**Fick's 2$^{nd}$ Law of Diffusion Fitting:** All code as written in Python using the Spyder integrated development environment (IDE). Spyder belongs to the MIT License and is distributed through the Anaconda environment. The curve_fit() function with a variety of initial values and boundary conditions were used to verify fit robustness. Two co-authors independently wrote code with similar results to verify algorithm accuracy.

**Density Functional Theory (DFT) Calculations:** All calculations were carried out using the generalized gradient approximation (GGA) of Perdew, Burke, and Ernzerhof (PBE)[54] using the projector augmented wave method (PAW)[55,56] as implemented in VASP[57] Lattice constant optimization for single quintuple layers (1QL) of Bi$_2$Se$_3$, Bi$_2$Se$_2$O, Bi$_2$SeO$_2$, and Bi$_2$O$_3$, all in the $R\bar{3}m$ structure, were performed with a 12x12x1 k-point mesh and energy cutoff of 500 eV, and were converged to within 1x10$^{-8}$ eV. The bismuth potentials used included 23 valence electrons, while the selenium and oxygen potentials used each included 6 valence electrons. A van der Waals correction was added using the DFT-D3 method of Grimme.[58] By relaxing the 1QLs at a variety of lattice constants around an expected minimum value, an energy vs. lattice constant curve was generated, which was parabolic as expected in the elastic limit of small deformations away from equilibrium. Optimized lattice constants were identified by finding the lattice constant for which this parabolic energy function is at a minimum for each material. Forces in the optimized structures were converged to ≤ 3 meV/Angstrom in each 1QL.

The electronic calculations for Bi$_2$Se$_3$, Bi$_2$O$_3$, and WSe$_2$ used in generating the bands and band offsets presented in the supplement used the same settings as above, but spin-orbit coupling was included. The tungsten potential used in the WSe$_2$ calculation included 14 valence electrons.

The Bi$_2$Se$_3$/WSe$_2$ heterostructure calculations presented in the supplement were performed using a 0° aligned structure with 16 unit cells of Bi$_2$Se$_3$ and 25 unit cells of WSe$_2$. Each layer was under about 0.57% compressive and tensile strain, respectively. All calculations used a bismuth potential that includes 5 valence electrons and selenium and tungsten potentials that each include 6 valence electrons. The energy cutoff was 500 eV, and van der Waals corrections were again included using the DFT-D3 method of Grimme.[58] The k-point mesh for relaxation and self-consistent calculations was 3x3x1. The forces in the relaxed structure were converged to less than 3 meV/Angstrom. Electronic calculations included spin-orbit coupling, and energies were converged to within 1x10$^{-7}$ eV.

The phonon band structures presented in the supplement were based on 2x2 tiled supercells of 1QL of each material: Bi$_2$Se$_3$, Bi$_2$Se$_2$O, Bi$_2$SeO$_2$, and Bi$_2$O$_3$, all in the $R\bar{3}m$ structure. A Density Functional Perturbation Theory (DFPT) calculation was carried out in VASP[57] using a 3x3x1 k-point mesh, and phonon band structures were generated using Phonopy.[59] Neither spin-orbit coupling nor a van der Waals correction were included in these calculations since only general questions about stability at 0K were addressed. Bismuth potentials used included 5 valence electrons, while selenium and oxygen potentials each included 6 valence electrons.

**Supporting Information**
STEM-EDS information (Section S1); Phonon Bands (Section S2); As-grown Bi$_2$Se$_3$-WSe$_2$ Theory Calculations (Section S3); WSe$_2$ PL spectra analysis and Lorentzian fitting (Section S4); Bi$_2$O$_3$ vs. Bi$_2$Se$_3$ on WSe$_2$ (Section S5); Control Experiments with monolayer WSe$_2$ (Section S6); Fick's 2$^{nd}$ Law of Diffusion Amplifying Discussion (Section S7); Long-term stability (Section S8); Very-Low power laser vacuum evolution (Section S9); Bi$_x$O$_y$Se$_z$ on monolayer WS$_2$ (Section S10); AFM scan of MBE grown Bi$_2$Se$_3$ (Section S11).

Permeability. *Chem. Commun.* **2019**, *55* (24), 3493–3496. https://doi.org/10.1039/C8CC10077B.

(14) Jiang, H.-Y.; Liu, J.; Cheng, K.; Sun, W.; Lin, J. Enhanced Visible Light Photocatalysis of Bi2O3 upon Fluorination. *J. Phys. Chem. C* **2013**, *117* (39), 20029–20036. https://doi.org/10.1021/jp406834d.

(15) Wang, J.; Neaton, J. B.; Zheng, H.; Nagarajan, V.; Ogale, S. B.; Liu, B.; Viehland, D.; Vaithyanathan, V.; Schlom, D. G.; Waghmare, U. V.; Spaldin, N. A.; Rabe, K. M.; Wuttig, M.; Ramesh, R. Epitaxial BiFeO3 Multiferroic Thin Film Heterostructures. *Science* **2003**. https://doi.org/10.1126/science.1080615.

(16) Shinde, P. V.; Shinde, N. M.; Shaikh, S. F.; Lee, D.; Yun, J. M.; Woo, L. J.; Al-Enizi, A. M.; Mane, R. S.; Kim, K. H. Room-Temperature Synthesis and CO2-Gas Sensitivity of Bismuth Oxide Nanosensors. *RSC Adv.* **2020**, *10* (29), 17217–17227. https://doi.org/10.1039/D0RA00801J.

(17) Cao, Y.; Fatemi, V.; Fang, S.; Watanabe, K.; Taniguchi, T.; Kaxiras, E.; Jarillo-Herrero, P. Unconventional Superconductivity in Magic-Angle Graphene Superlattices. *Nature* **2018**, *556* (7699), 43–50. https://doi.org/10.1038/nature26160.

(18) Liu, X.; Hersam, M. C. 2D Materials for Quantum Information Science. *Nat. Rev. Mater.* **2019**, *4* (10), 669–684. https://doi.org/10.1038/s41578-019-0136-x.

(19) Qiu, H.; Zhou, W.; Guo, W. Nanopores in Graphene and Other 2D Materials: A Decade's Journey toward Sequencing. *ACS Nano* **2021**, *15* (12), 18848–18864. https://doi.org/10.1021/acsnano.1c07960.

(20) Deng, D.; Novoselov, K. S.; Fu, Q.; Zheng, N.; Tian, Z.; Bao, X. Catalysis with Two-Dimensional Materials and Their Heterostructures. *Nat. Nanotechnol.* **2016**, *11* (3), 218–230. https://doi.org/10.1038/nnano.2015.340.

(21) Das, S.; Sebastian, A.; Pop, E.; McClellan, C. J.; Franklin, A. D.; Grasser, T.; Knobloch, T.; Illarionov, Y.; Penumatcha, A. V.; Appenzeller, J.; Chen, Z.; Zhu, W.; Asselberghs, I.; Li, L.-J.; Avci, U. E.; Bhat, N.; Anthopoulos, T. D.; Singh, R. Transistors Based on Two-Dimensional Materials for Future Integrated Circuits. *Nat. Electron.* **2021**, *4* (11), 786–799. https://doi.org/10.1038/s41928-021-00670-1.

(22) Bhat, A.; Anwer, S.; Bhat, K. S.; Mohideen, M. I. H.; Liao, K.; Qurashi, A. Prospects Challenges and Stability of 2D MXenes for Clean Energy Conversion and Storage Applications. *Npj 2D Mater. Appl.* **2021**, *5* (1), 1–21. https://doi.org/10.1038/s41699-021-00239-8.

(23) Fathi-Hafshejani, P.; Azam, N.; Wang, L.; Kuroda, M. A.; Hamilton, M. C.; Hasim, S.; Mahjouri-Samani, M. Two-Dimensional-Material-Based Field-Effect Transistor Biosensor for Detecting COVID-19 Virus (SARS-CoV-2). *ACS Nano* **2021**, *15* (7), 11461–11469. https://doi.org/10.1021/acsnano.1c01188.

(24) Rodriguez-Lamas, R.; Pirovano, C.; Stangl, A.; Pla, D.; Jónsson, R.; Rapenne, L.; Sarigiannidou, E.; Nuns, N.; Roussel, H.; Chaix-Pluchery, O.; Boudard, M.; Jiménez, C.; Vannier, R.-N.; Burriel, M. Epitaxial LaMnO3 Films with Remarkably Fast Oxygen Transport Properties at Low Temperature. *J. Mater. Chem. A* **2021**, *9* (21), 12721–12733. https://doi.org/10.1039/D0TA12253J.

(25) Stilhano Vilas Boas, C. R.; Sturm, J. M.; Milov, I.; Phadke, P.; Bijkerk, F. Room Temperature Oxygen Exchange and Diffusion in Nanometer-Thick ZrO2 and MoO3 Films. *Appl. Surf. Sci.* **2021**, *550*, 149384. https://doi.org/10.1016/j.apsusc.2021.149384.
21

# Supporting Information

**Room-temperature oxygen transport in nano-thin $Bi_xO_ySe_z$ enables precision modulation of 2D materials**

Zachariah Hennighausen[1,*], Bethany M. Hudak[2], Madeleine Phillips[2], Jisoo Moon[1], Kathleen M. McCreary[2], Hsun-Jen Chuang[3], Matthew R. Rosenberger[4], Berend T. Jonker[2], Connie H. Li,[2] Rhonda M. Stroud[2], and Olaf M. van 't Erve[2,*]

[1] NRC Postdoc Residing at the Materials Science and Technology Division, United States Naval Research Laboratory, Washington, D.C. 20375, USA
[2] Materials Science and Technology Division, United States Naval Research Laboratory, Washington, D.C. 20375, USA
[3] Nova Research, Inc., Alexandria, VA 22308, USA
[4] University of Notre Dame, Notre Dame, IN 46556, USA

# Section S1. STEM-EDS quantification of $Bi_xO_ySe_z$ vs. $Bi_2Se_3$

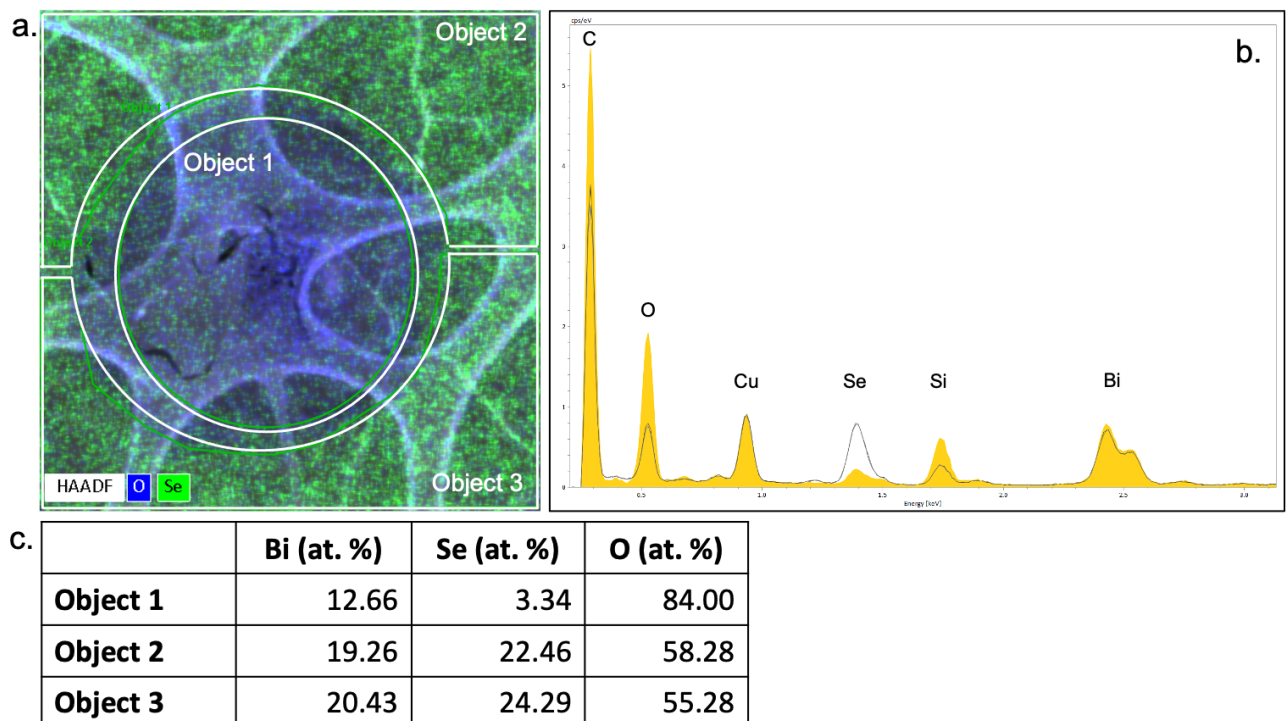

| | Bi (at. %) | Se (at. %) | O (at. %) |
|---|---|---|---|
| **Object 1** | 12.66 | 3.34 | 84.00 |
| **Object 2** | 19.26 | 22.46 | 58.28 |
| **Object 3** | 20.43 | 24.29 | 55.28 |

**Figure S1. STEM-EDS quantification.** (a) STEM-EDS map with object regions drawn in Bruker Esprit 2.0 software to quantify the chemical change between unaltered and laser-oxygen exposed $Bi_2Se_3$. For readability, the approximate positions of the objects has been added. (b) EDS spectra of Object 1 (yellow), Object 2 (black line), and Object 3 (grey line), normalized to the Cu system peak. (c) Chart displaying atomic percentage of Bi, Se, and O for each object. The values reported in the main text are taken from this quantification. Values in Objects 2 and 3 were averaged together to obtain the pristine $Bi_2Se_3$ values. Cu is a system peak. Carbon oxide contamination is common on TEM samples, and therefore some O is present on the pristine $Bi_2Se_3$ (Objects 2 and 3).

The SAED data combined with the STEM-EDS data strongly indicates that the laser-oxygen exposure of the $Bi_2Se_3$ is removing Se atoms and replacing them with O atoms, resulting in the production of a $Bi_xO_ySe_z$ compound with a reduced lattice constant (3.65 Å), which is quite close to $R\bar{3}m$ $Bi_2O_3$ (3.642 Å).

The shape and size of this spot corresponds well with the laser spot size, which is approximately circular with a Gaussian energy distribution. A great majority of the energy is confined within a ~800nm diameter area. The laser spot size skewed when focused on the TEM grid because the grids contains significantly more wrinkles compared to $SiO_2$, which causes areas to vary in focus and the laser power density applied.

## Section S2. Phonon bands for various bulk Bi$_2$O$_y$Se$_z$ from Bi$_2$Se$_3$ to Bi$_2$O$_3$

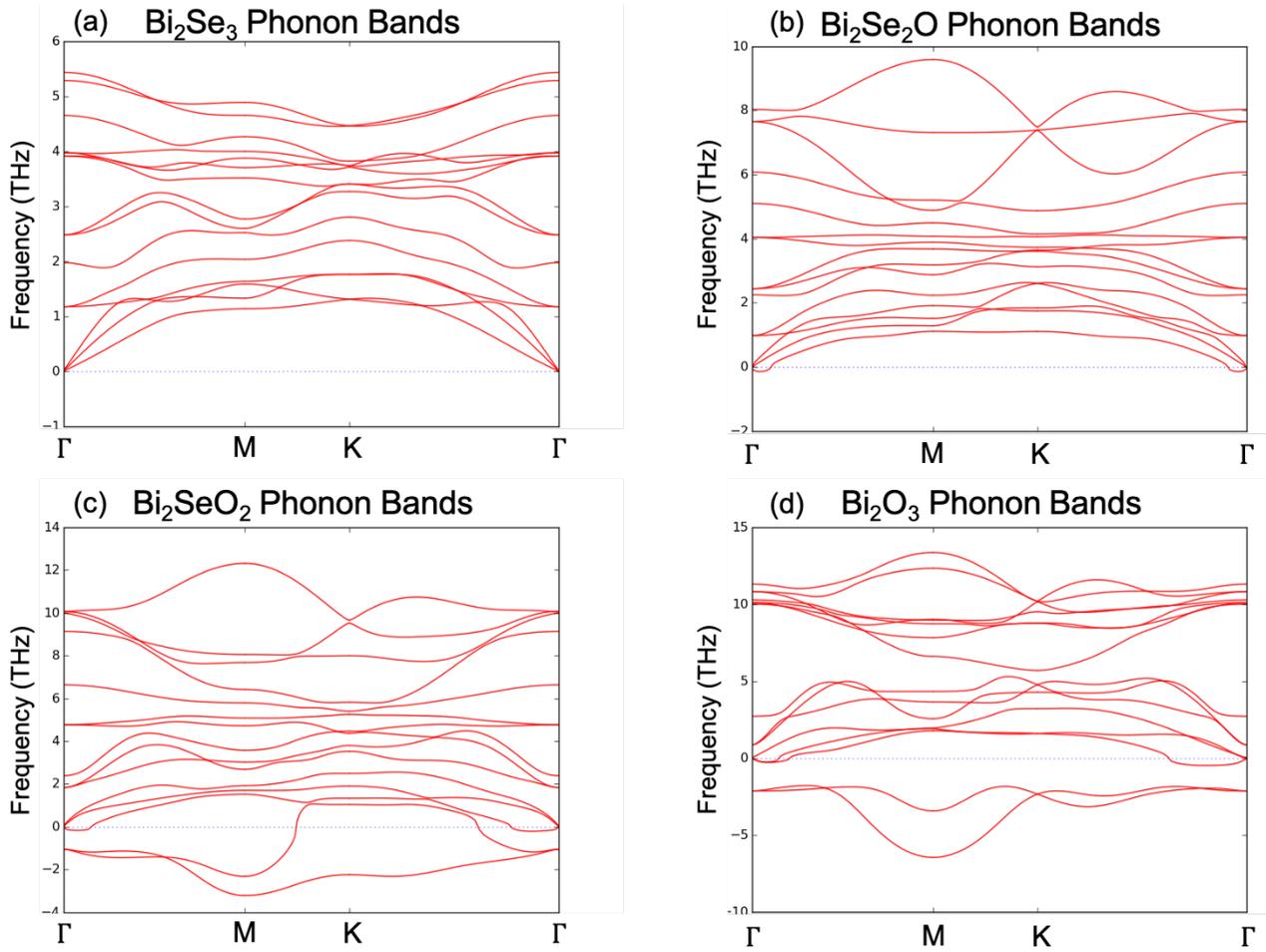

**Figure S2.** Phonon band structures for (a) Bi$_2$Se$_3$ (b) Bi$_2$Se$_2$O (c) Bi$_2$SeO$_2$ and (d) Bi$_2$O$_3$, all with the $R\bar{3}m$ structure. Each set of phonon bands was computed using a 5-atom unit tiled 2x2x1 using the Phonopy program.[1] Density functional perturbation theory was carried out using a 3x3x1 k-point mesh, without spin-orbit coupling. The monolayer of Bi$_2$Se$_3$ is stable at 0K, as expected. Bi$_2$Se$_2$O may also be considered stable at 0K, since the imaginary phonon mode is not at a high symmetry point, and thus not well sampled by the k-point mesh chosen. Bi$_2$SeO$_2$ and Bi$_2$O$_3$ are not found to be stable at 0K. However, such materials might be stabilized through mechanisms present in experiment, but not considered in the calculations. For example, other Bi$_2$O$_3$ phases, such as δ-Bi$_2$O$_3$, are only stable above certain temperatures or when coupled to certain substrates.[2] Additionally, rhombohedral Bi$_2$O$_3$ has been previously reported in bulk form when stabilized with dopants or a substrate, raising the possibility that a combination of structural, chemical, and environmental factors is stabilizing this rare phase of Bi$_2$O$_3$ in our experiments.[3–5] See main text for further discussion.

## Section S3. As-grown $Bi_2Se_3$-$WSe_2$ Theory Calculations

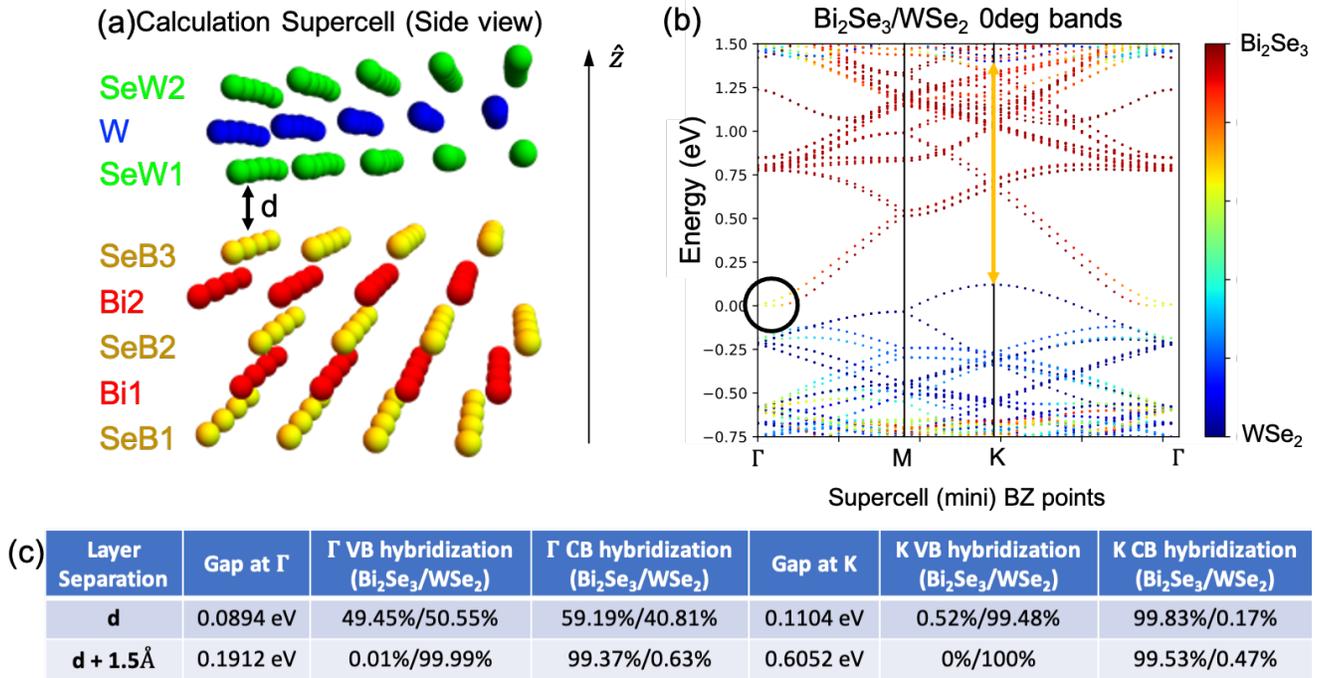

| Layer Separation | Gap at Γ | Γ VB hybridization ($Bi_2Se_3$/$WSe_2$) | Γ CB hybridization ($Bi_2Se_3$/$WSe_2$) | Gap at K | K VB hybridization ($Bi_2Se_3$/$WSe_2$) | K CB hybridization ($Bi_2Se_3$/$WSe_2$) |
|---|---|---|---|---|---|---|
| d | 0.0894 eV | 49.45%/50.55% | 59.19%/40.81% | 0.1104 eV | 0.52%/99.48% | 99.83%/0.17% |
| d + 1.5Å | 0.1912 eV | 0.01%/99.99% | 99.37%/0.63% | 0.6052 eV | 0%/100% | 99.53%/0.47% |

**Figure S3. Photoluminescence (PL) quenching in $Bi_2Se_3$/$WSe_2$ bilayer.** (a) Side view of 0° twisted supercell of $Bi_2Se_3$/$WSe_2$. $WSe_2$ layer is under ~0.57% tensile strain, and $Bi_2Se_3$ layer is under ~0.57% compressive strain. Relaxed equilibrium distance between layers is d=3.276 Å. (b) Band structure along high symmetry $\Gamma$-M-K-$\Gamma$ line. Band points corresponding to states localized entirely on $WSe_2$ layer are dark blue. Band points corresponding to states localized on $Bi_2Se_3$ layer are dark red. Band points corresponding to states with weight on both layers are colored according to the scale bar at the far right. States corresponding to the conduction band (CB) minimum at Gamma have roughly equal weight in each layer. The transition associated with the A exciton in $WSe_2$ is indicated by the orange double-headed arrow. (c) Table shows band gap sizes and band edge hybridizations for the heterostructure at the equilibrium separation, d, and at a separation d + 1.5 Å, simulating the effect of oxygen intercalation between layers. PL quenching in the pristine heterostructure is attributed to the hybridized states at the Gamma point presenting a non-radiative decay pathway for carriers excited at K in $WSe_2$. Increasing the interlayer separation causes the states at Gamma to be well layer-polarized again, removing the non-radiative decay pathway and restoring the PL in $WSe_2$. Similar behavior is expected for the $Bi_2O_3$/$WSe_2$ monolayer. (See Figure S7).
.

# Section S4. Photoluminescence (PL) spectra decomposition into Lorentzian functions, and analysis of individual Lorentzian evolution due to laser-oxygen exposure

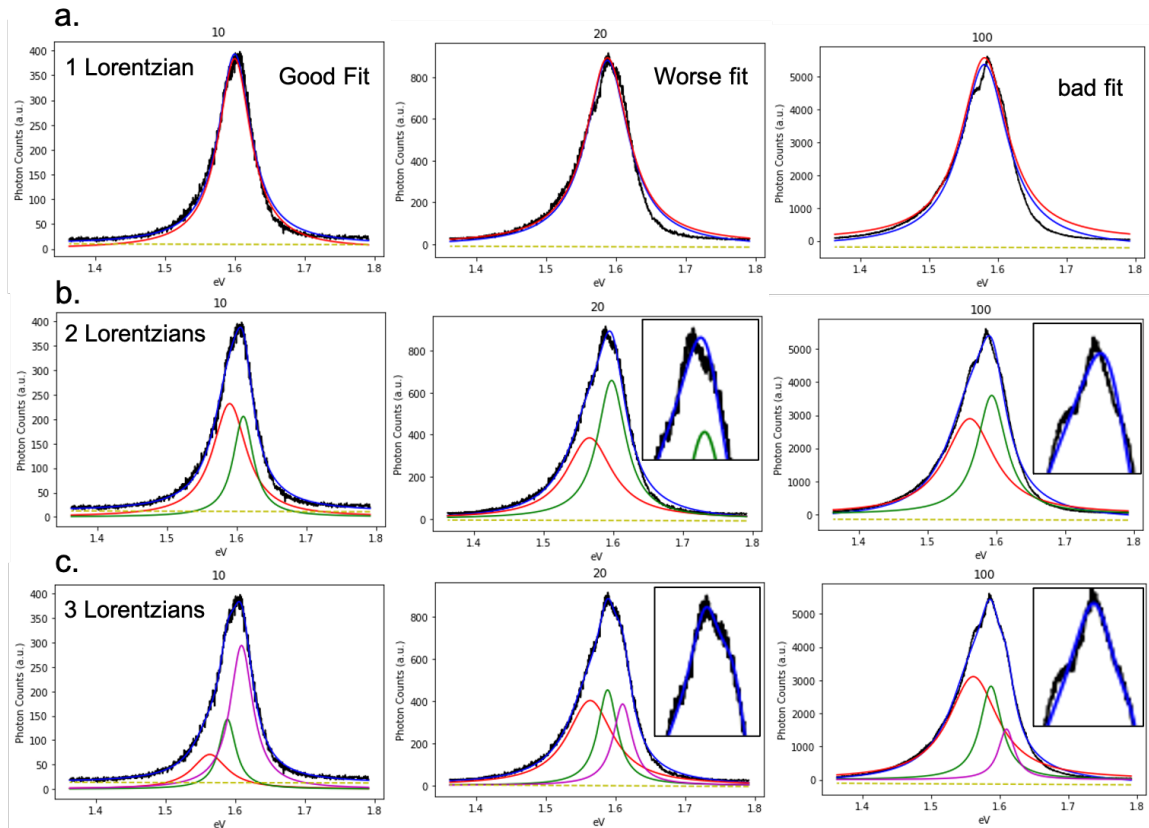

**Figure S4. Multi-Lorentzian function fitting reveals numerous radiative pathways compose $Bi_xO_ySe_z$-$WSe_2$ 2D heterostructure spectra.** Three different curve decomposition functions were applied using a Linear background and multiple Lorentzians. (a) Single Lorentzian with a linear background show a good fit initially, but evolve to bad fits as the spectra shape changes. (b) Two Lorentzians with a linear background fit well throughout with the exception of the peak position, where a small deviation is observed. (c) Three Lorentzians with a linear background fit well throughout. Further, the higher intensity spectra qualitatively contain three features (e.g., sharp peak, and two shoulders), suggesting three effects are present (e.g., exciton species, exciton-phonon scattering induced phonon sidebands). Spectra above correspond to the PL evolution plots shown in Figure 6d-e. Expanded representative fitting is shown in Figure S6.

The $Bi_xO_ySe_z$-$WSe_2$ spectra intensity, peak position, and shape evolve during laser-air exposure. The spectral line shape from a single radiative recombination is frequently described using a Lorentzian function. In our system, the spectra shape often deviates from a single Lorentzian, suggesting multiple radiative recombination pathways and effects are present. For example, Figure S4 shows spectra labeled "100" (right side) where a sharp peak and two shoulders are visible, strongly deviating from a single Lorentzian fit. The PL spectra in monolayer TMDs are frequently composed of numerous radiative recombination pathways or effects. For example, monolayer $WSe_2$ contains a host of excitons at different peak positions that modulate intensity with external gating,[6] and phonon sidebands due to exciton-phonon scattering shift the spectra shape.[7] The evolving interaction between the materials during laser-oxygen exposure likely alters the exciton recombination pathways and surrounding dielectric environment, thereby changing the relative strength of different quasiparticle species and scattering processes (e.g., excitons, exciton-phonon scattering). Below we compare the differences in Two-Lorentzian vs. Three-Lorentzian fitting, making the case that Three-Lorentzian fitting better captures the system.

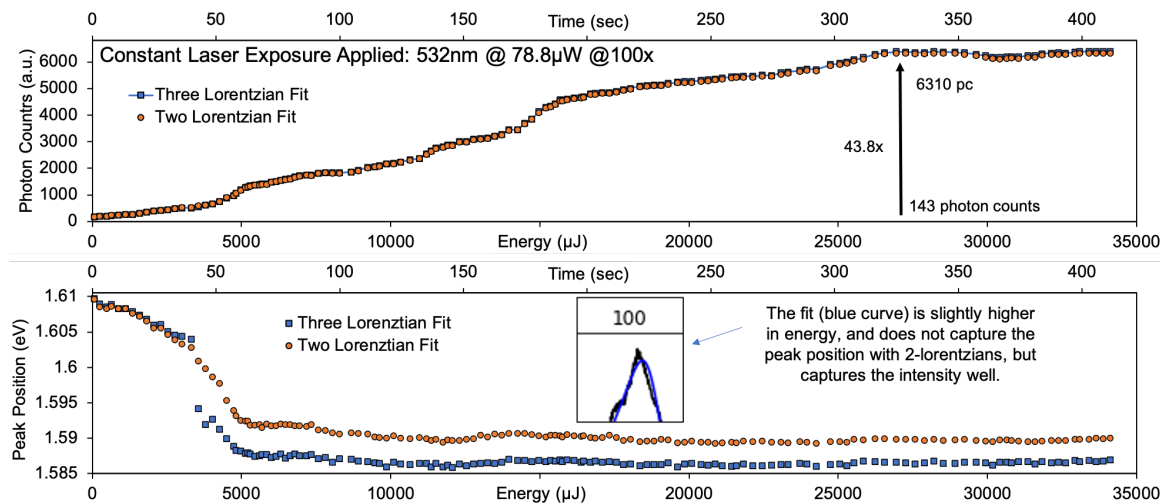

**Figure S5. Comparing PL intensity and peak position evolution by applying Three-Lorentzian vs. Two-Lorentzian fitting.** While the PL intensity evolution (top) shows equivalent behavior, the PL peak position evolution (bottom) shows a difference. The difference in peak position is likely because the Two-Lorentzian captures the peak less accurately than the Three-Lorentzian fitting (inset).

As shown in Figure S5, Two-Lorentzian vs. Three-Lorentzian fitting produce equivalent results for the PL intensity evolution, but diverge slightly for the Three-Lorentzian fitting. The difference in peak position is likely because the Two-Lorentzian captures the peak less accurately than the Three-Lorentzian fitting. Figure S4 shows that a single Lorentzian with linear background fitting is insufficient, in agreement with qualitative analysis of the spectra, where some contain a sharp peak and two shoulders, suggesting multiple radiative recombination pathways are present. The evolution of the three Lorentzian functions, as well as representative fitting, are shown next.

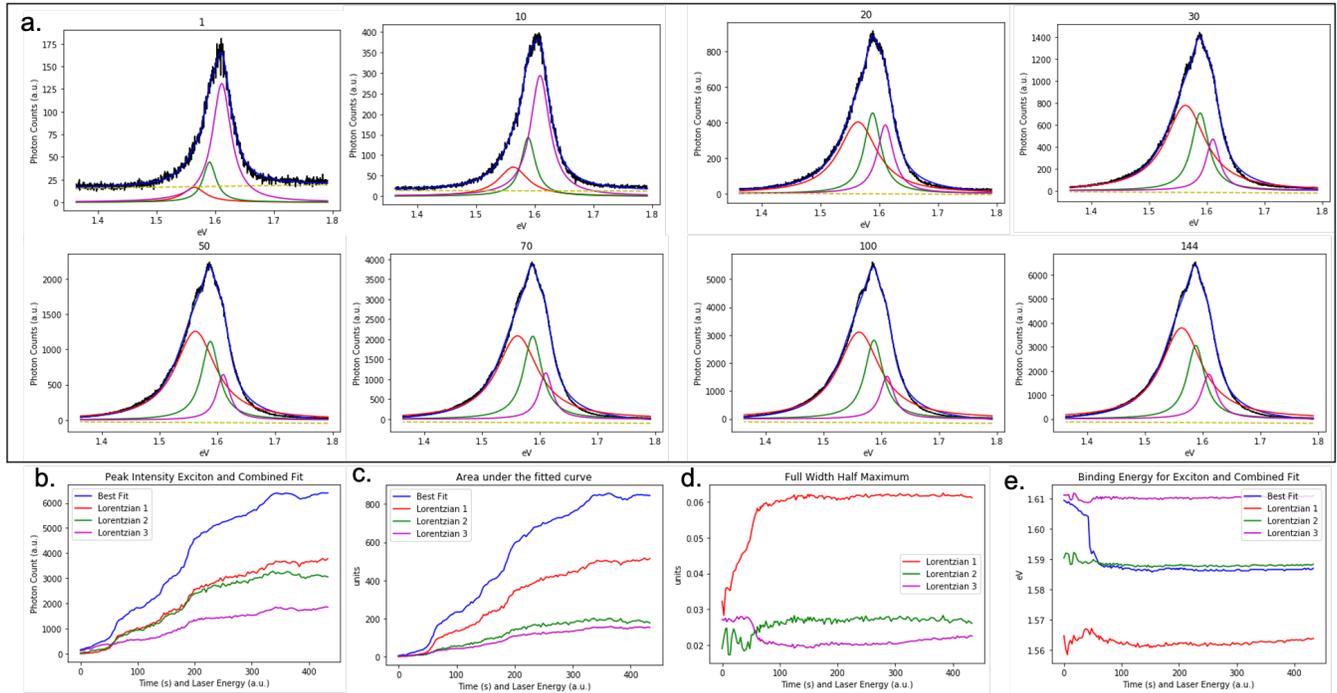

**Figure S6. Bi$_x$O$_y$Se$_2$-WSe$_2$ 2D heterostructure PL evolution under laser-oxygen exposure showing Three-Lorentzian fitting with a linear background.** (a)Representative fitting and (b)-(e) Lorentzian evolution for the data shown in Figure 6c-e. (a) Spectra at regular intervals to demonstrate the good fit throughout. (b) peak intensity. (c) Area under the curve. (d) Full-width-half-maximum (FWHM). (e) Peak position.

Figure S6 shows the Lorentzian functions evolving relative intensity, area, FWHM, and peak position throughout laser exposure. A great majority of the changes happen before the ~65s mark, suggesting the exciton recombination pathways and scattering are most influenced as oxygen begins to intercalate. The peak position rapidly decreases before plateauing, all the while the intensity keeps increasing, suggesting at least one mechanism that asymmetrically targets the peak position switches from active to dormant around the 65s mark. The data corresponds to Figure 6c-e.

## Section S5. 1QL Bi$_2$O$_3$ vs. 1QL Bi$_2$Se$_3$ band structures, and 1L WSe$_2$ energy band alignment

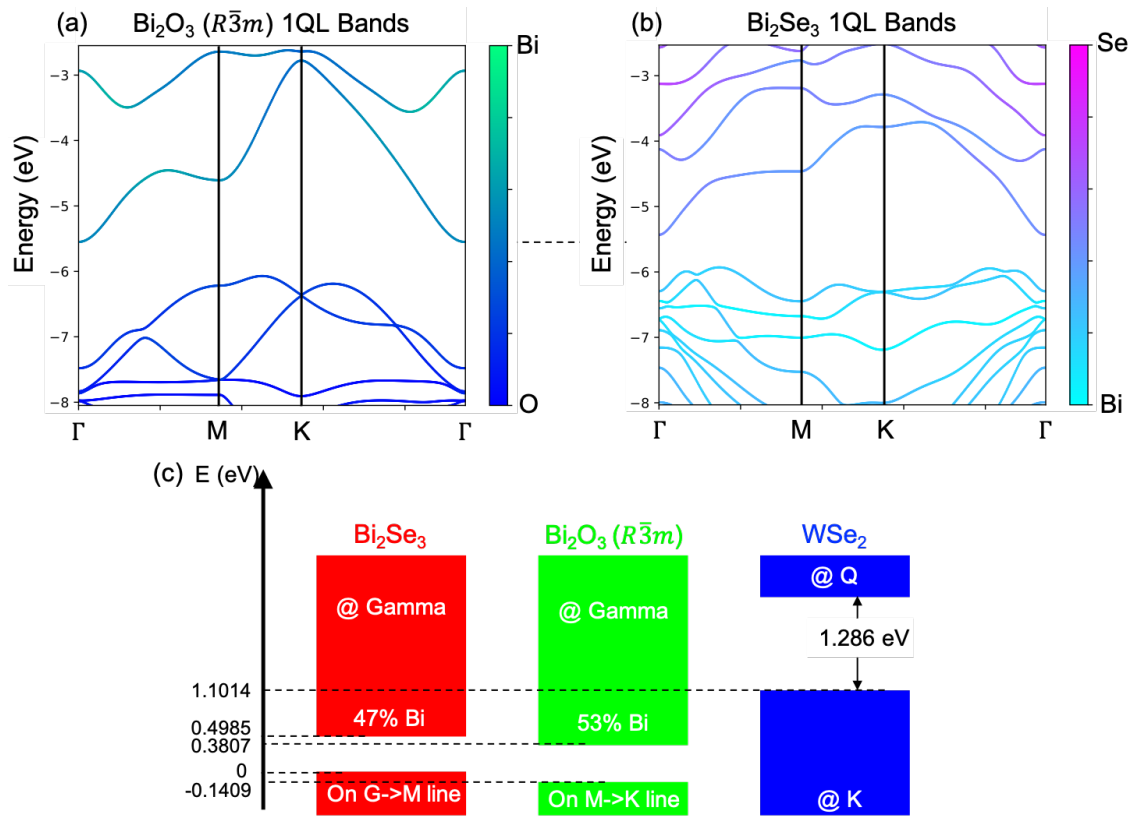

**Figure S7. (a)** Band structure of 1 quintuple layer (1QL) of Bi$_2$O$_3$ in the $R\bar{3}m$ structure. **(b)** Band structure of 1 quintuple layer of Bi$_2$Se$_3$ **(c)** Band alignments for Bi$_2$Se$_3$, Bi$_2$O$_3$, and WSe$_2$. The lowest conduction bands in the Bi$_2$Se$_3$ and Bi$_2$O$_3$ are similar in shape and Bi weight at Gamma. The biggest single contribution to the weight at the Gamma point is in Bi p$_z$ orbitals for both Bi$_2$Se$_3$ and Bi$_2$O$_3$. The band alignments with WSe$_2$ are also very similar. This suggests that the hybridization behavior of a Bi$_2$O$_3$/WSe$_2$ heterostructure may be similar to the behavior computed in Section S3 for Bi$_2$Se$_3$/WSe$_2$.

# Section S6. Laser-air exposure of monolayer WSe$_2$ shows comparatively small changes

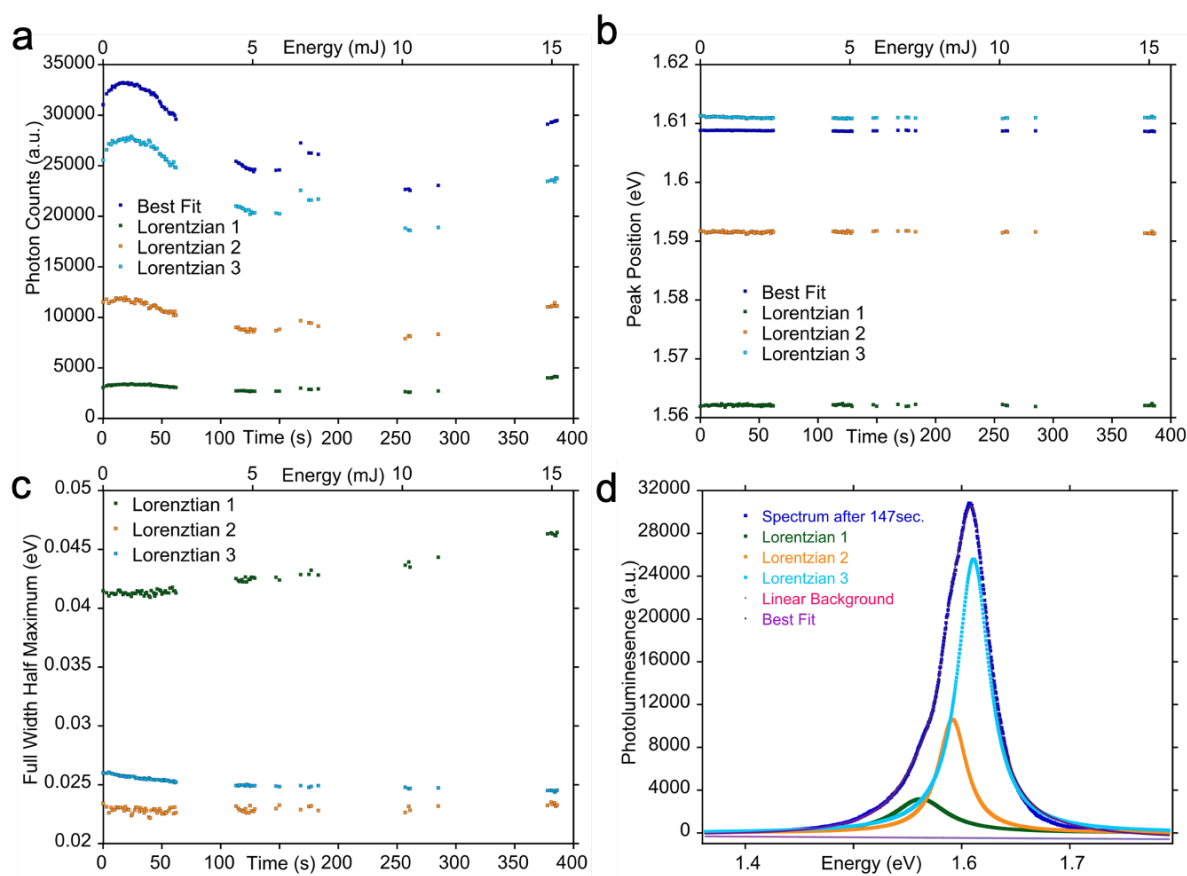

**Figure S8. Monolayer WSe$_2$ (without Bi$_x$O$_y$Se$_z$) subject to laser-oxygen exposure.** (a)-(c) show the evolution of PL (a) intensity, (b) peak position and (c) FWHM of monolayer WSe$_2$. Although changes are detected, the changes are comparatively negligible, and do not qualitatively follow the same behavior. For example, the PL intensity increases only 6%, before decreasing, while the peak position remains nearly constant, a clear deviation from Bi$_x$O$_y$Se$_z$-WSe$_2$ 2D heterostructures. (d) Representative PL spectra with Lorentzian and linear background fitting.

No significant or notable changes are observed when as-grown monolayer WSe$_2$ is subject to laser-oxygen exposure, demonstrating that PL modification in Bi$_x$O$_y$Se$_z$-WSe$_2$ 2D heterostructures requires Bi$_x$O$_y$Se$_z$, suggesting no prominent chemical modification of WSe$_2$.

# Section S7. Amplifying Discussion for Fick's 2nd Law of Diffusion with Fixed Boundary Conditions

Consider Fick's 2nd Law of Diffusion equation in one dimension, where $D$ is the diffusion coefficient, $C$ is the concentration, $x$ is distance, and $t$ is time.

$$\frac{\partial C}{\partial t} = D \frac{\partial^2 C}{\partial x^2} \tag{1}$$

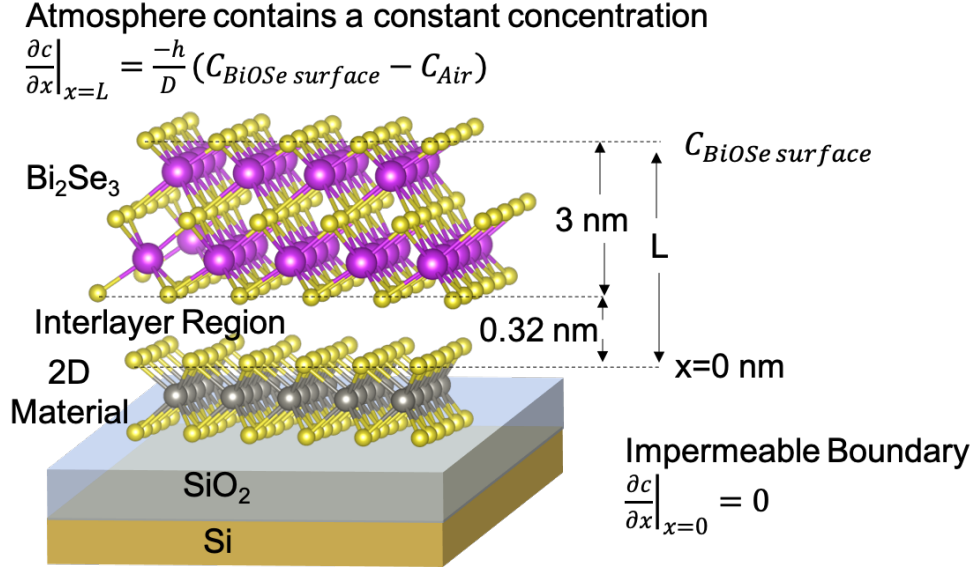

Figure S9. Schematic defining terms for amplifying discussion of Fick's 2nd Law of Diffusion.

We define the following boundary conditions. The impermeable boundary condition is defined as when no flux happens across the boundary, which is at x=0 nm. We define a constant concentration in the air; however, we allow for a finite mass transfer coefficient ($h$), which acts as an energy and absorption barrier between the air and $Bi_xO_ySe_z$.

$$\left.\frac{\partial c}{\partial x}\right|_{x=0} = 0 \tag{2}$$

$$\left.\frac{\partial c}{\partial x}\right|_{x=L} = \frac{-h}{D}(C_{BiOSe\ surface} - C_{Air}) \tag{3}$$

$$c(x, t=0) = c_{initial} \tag{4}$$

We redefine the variables to become dimensionless. $K$ is the partition coefficient.

$$\theta = \frac{C - K*C_{Air}}{C_{initial} - K*C_{Air}} \tag{5}$$

$$x^* = \frac{x}{L} \tag{6}$$

$$F = \frac{Dt}{L^2} \tag{7}$$

Substitute equations (5), (6), and (7) into Eq. (1).

$$\left(\frac{D}{L^2}\frac{\partial t}{\partial F}\right)\frac{\partial C}{\partial t} = D\frac{\partial^2 C}{\partial x^2}\left(\frac{\partial^2 \theta}{\partial C^2}\right)\left(\frac{\partial^2 x}{\partial x^{*2}}\frac{1}{L^2}\right) \implies \frac{\partial \theta}{\partial F} = \frac{\partial^2 \theta}{\partial x^{*2}} \tag{8}$$

We solve Eq. (8) using the technique *Separation of Variables* while applying the boundary conditions defined in (2), (3), and (4).

$$c(x,t) = c_{air} - (c_{air} - c_{Initial})\sum_{n=1}^{\infty} C_n \exp(-\zeta_n^2 F)\cos\left(\zeta_n \frac{x}{L}\right) \tag{9}$$

$$C_n = \frac{4\sin(\zeta_n)}{2\zeta_n + \sin(2\zeta_n)}; \quad \zeta_n \tan(\zeta_n) = B; \quad B = \frac{hL}{KD}$$

Table 1. Measured diffusion constants and $h/K$ values for the $Bi_xO_ySe_z$-$WSe_2$ 2D heterostructure shown in Figure 7.

|  | Diffusion Constant (m²s⁻¹) | h/K (m/s) |
|---|---|---|
| 1st Vacuum | 4.92E-19 | 1.26E-10 |
| 2nd Air | 7.65E-19 | 2.28E-10 |
| 2nd Vacuum | 6.16E-19 | 1.59E-10 |
| Part I: Low Pressure $O_2$ | 4.49E-19 | 1.09E-10 |
| Part II: Full Pressure $O_2$ | 1.32E-18 | 3.29E-10 |
| 99.9% $N_2$ | 4.155E-19 | 1.03E-10 |
| 99.99% $O_2$ | 2.613E-18 | 5.71E-10 |
| Low Power Laser (6.26µW) | 1.59E-19 | 4.44E-11 |
| 3rd Air | 6.57E-19 | 1.67E-10 |

The solution makes assumptions that deviate from experiment. We discuss the assumptions and how they might affect the measured value.

(1) We assume that the PL intensity is proportional to the oxygen concentration in the interlayer region. However, it is possible that a disproportionally small or large amount of oxygen is required to decouple the $Bi_xO_ySe_z$ and monolayer material. Layered materials have demonstrated the ability to store materials in ultra-dense configurations, raising the possibility that much larger amounts of oxygen are absorbed than we are assuming.[8] If a larger amount of oxygen is required to decouple the materials, the diffusion constant we measure is an underestimate.

(2) We assume at steady state the oxygen concentration is close to the concentration in air. However, it is possible that a different concentration of oxygen is absorbed at steady state. Numerous materials have demonstrated that they are able to absorb and store oxygen at greater densities than in air.[9] If a larger amount of oxygen is absorbed, the diffusion constant we measure is an underestimate.

(3) We assume the diffusion constant is remains constant throughout $Bi_xO_ySe_z$ and the interlayer region. We make this assumption because measuring the diffusion constant for the interlayer region is beyond the scope of this manuscript, and the DFT-predicted interlayer region (0.32nm) is only ~10% of the $Bi_xO_ySe_z$ thickness (3nm).

(4) We assume the oxygen does not diffuse through the 2D material. If oxygen diffuses into the 2D material, this would lead to an underestimation of the diffusion constant.

(5) We assume the oxygen does not react with the 2D material. It is possible the oxygen reacts with vacancy sites or partially diffuses into the 2D material. This would affect the PL intensity and amount of oxygen absorbed. If the reaction at vacancy sites encourages PL brightening, this would lead to an overestimation of the diffusion constant.

# Section S8. Long-term stable in N₂ atmosphere for 221 days

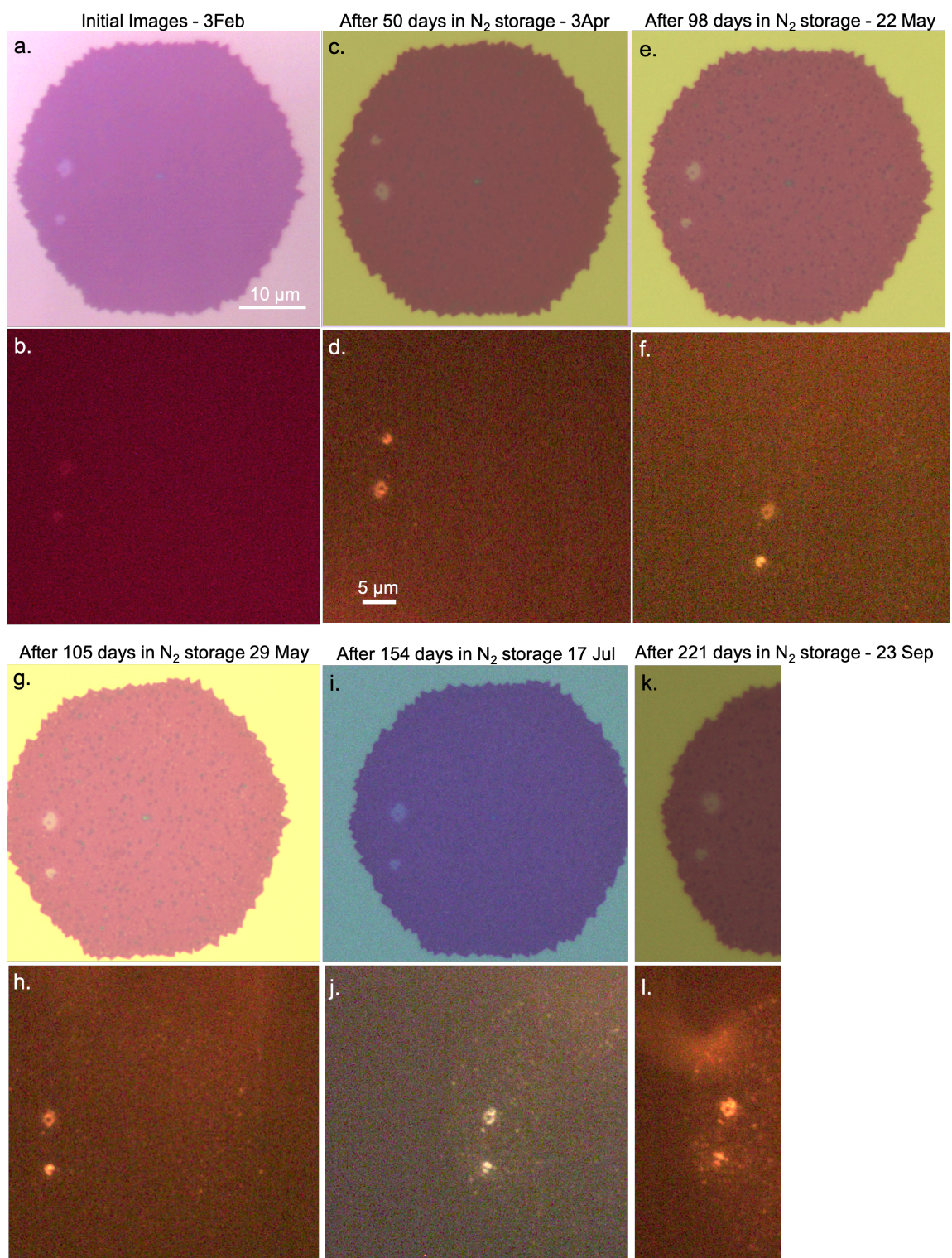

**Figure S10. Oxygen intercalated Bi$_x$O$_y$Se$_z$-WSe$_2$ 2D heterostructure is stable for mor than 221 days.** (a)-(i) Optical and fluorescence images of a Bi$_x$O$_y$Se$_z$-WSe$_2$ 2D heterostructure intercalated with oxygen at two distinct locations. (a)-(b) A different fluorescence setup was used for the initial measurement, giving the appearance of a diminished PL.

# Section S9. Very low power laser power in vacuum deintercalation

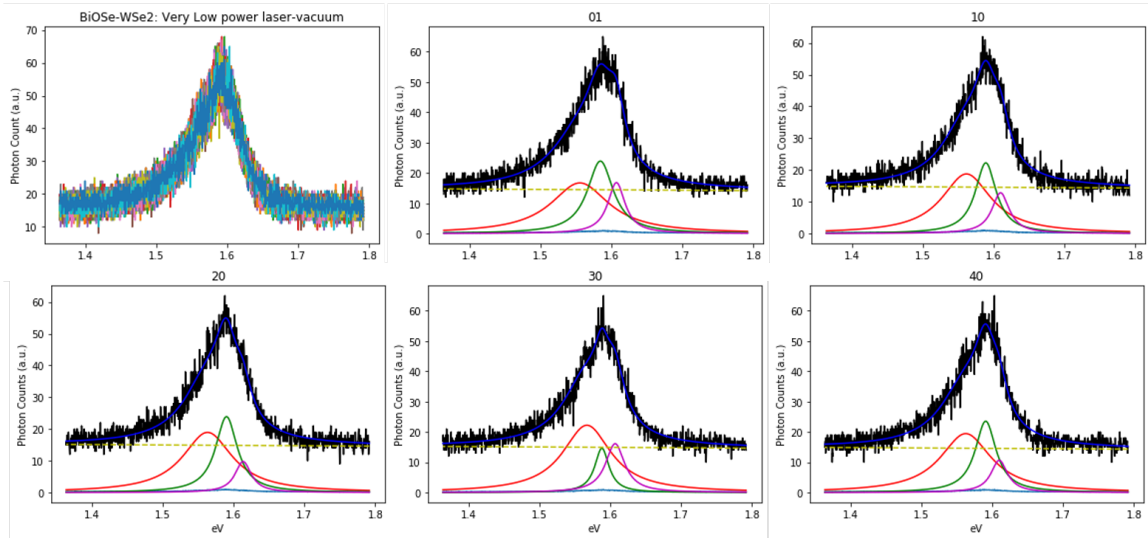

**Figure S11. Bi$_x$O$_y$Se$_z$-WS$_2$ 2D heterostructure subject to very low power laser-vacuum exposure.** The above spectra correspond to data shown in Figure 7. (Upper left) All 41 spectra plotted overlapping, qualitatively demonstrating little change observed over 88sec. (Remaining panels) PL spectrum with fitting (see Section S4). Although the data is noisier than PL spectra collected at low-power and standard power, it is well above the noise floor.

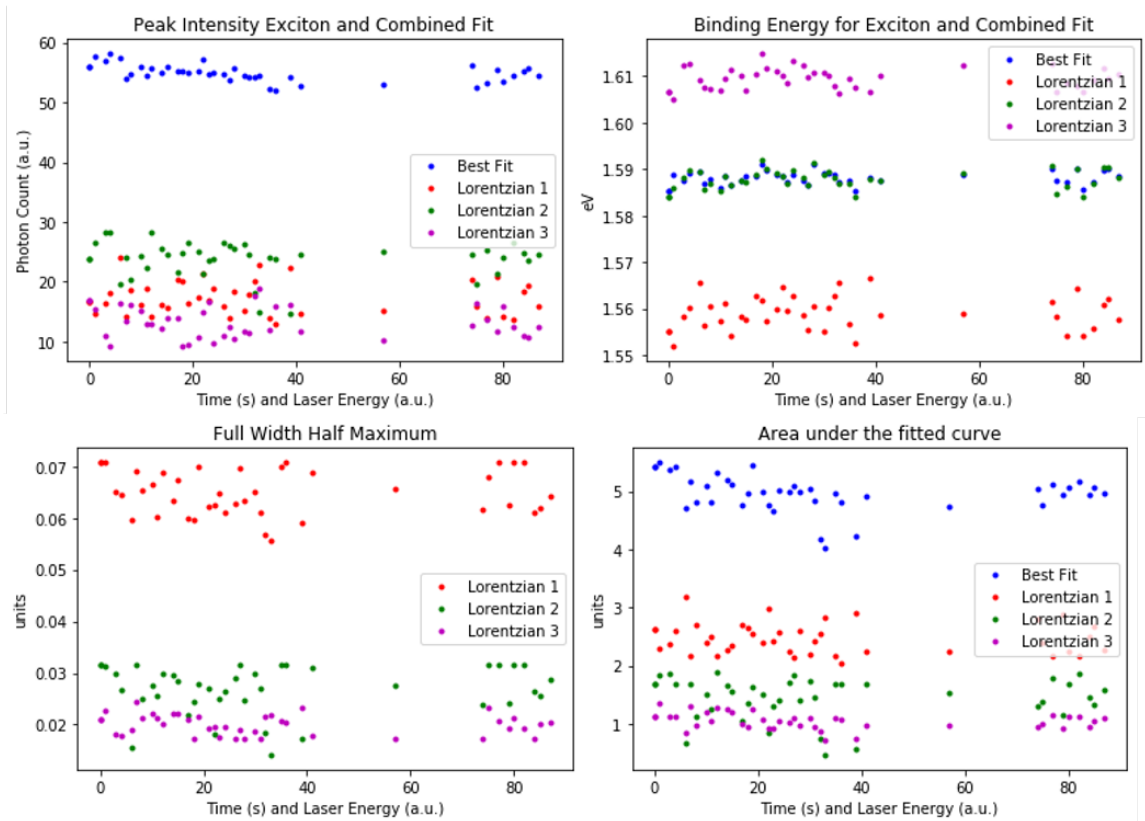

**Figure S12. Very low power laser-vacuum exposure evolution.** The above spectra correspond to data shown in Figure 7. Despite the length of exposure (88sec.) and number of spectra (41), no significant changes to the PL intensity, peak position, FWHM, and area under the curve are observed.

# Section S10. $Bi_xO_ySe_z$ can be applied to other 2D materials: $Bi_xO_ySe_z$-<u>WS$_2$</u> 2D Heterostructure

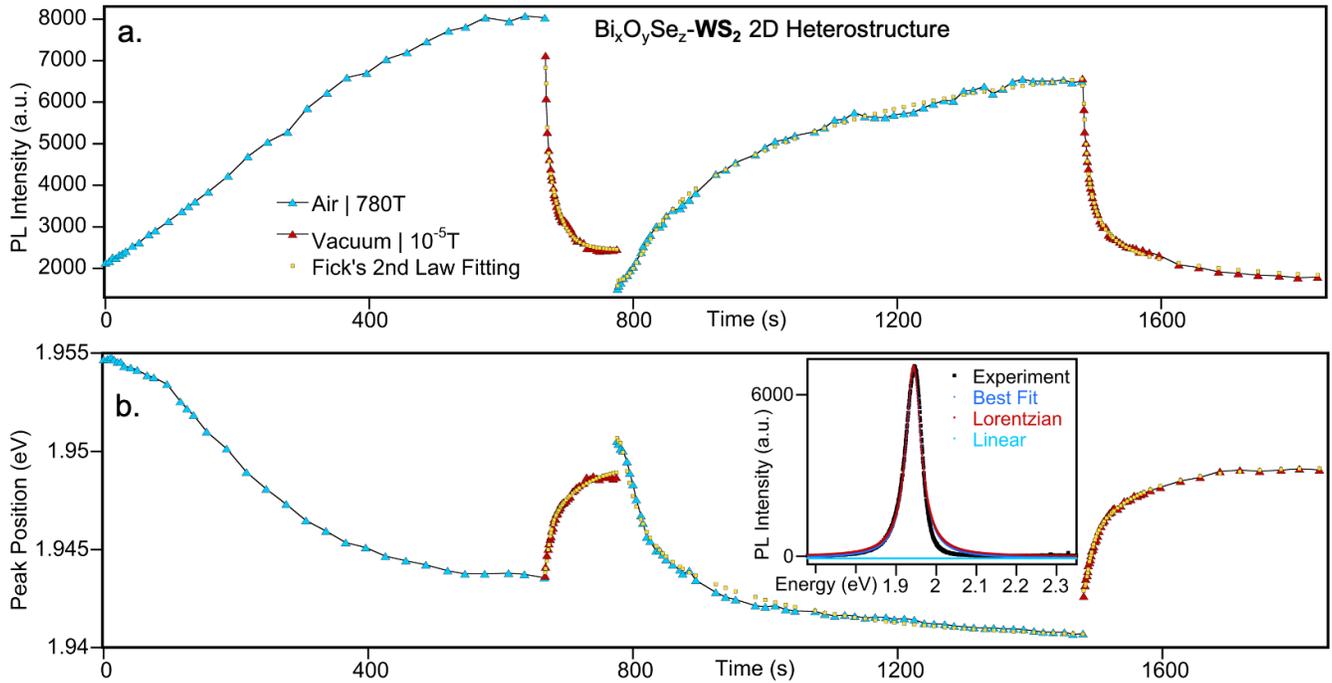

**Figure S13. $Bi_xO_ySe_z$-WS$_2$ 2D heterostructure subject to laser-air exposure.** Evolution of PL (above) intensity and (below) peak position. All subsequent exposures follow Fick's 2$^{nd}$ Law of Diffusion, in agreement with the $Bi_xO_ySe_z$-WSe$_2$ results.

Figure S12 shows that the properties of $Bi_xO_ySe_z$-WS$_2$ can be modulated, demonstrating that the effect is not confined to WSe$_2$ alone, raising the possibility it can be applied to other 2D materials. We believe that $Bi_xO_ySe_z$ can be applied to numerous other 2D materials to controllably modulate their properties with high spatial resolution, raising the possibility for a host of technologies, including spatially-selective and tunable graphene, long-lived interlayer excitons,[10] magnetism,[11] and ferroelectricity.[12]

Although previous work demonstrated that the PL of various TMD (i.e., WS$_2$, MoS$_2$, MoSe$_2$, and MoSe$_{(2-2x)}$S$_{2x}$) with Bi$_2$Se$_3$ 2D heterostructures could be modulated upward using laser-air exposure, none identified the effect as due to $Bi_xO_ySe_z$ and a rare $R\bar{3}m$ Bi$_2$O$_3$ phase.[13–15] One of the publications demonstrated that oxygen was central using Bi$_2$Se$_3$-MoS$_2$ 2D heterostructures, although the possibility of a chemical modification of Bi$_2$Se$_3$ was considered unlikely.

|  | 1st Vacuum | | 2nd Air | | 2nd Vacuum | |
|---|---|---|---|---|---|---|
|  | Intensity | Peak Position | Intensity | Peak Position | Intensity | Peak Position |
| Diffusion Coefficient (m$^2$s$^{-1}$) | 5.62E-19 | 3.77E-19 | 3.89E-20 | 1.15E-19 | 7.64E-19 | 2.29E-19 |
| h/k (m/s) | 9.41E-11 | 7.94E-11 | 9.69E-12 | 2.04E-11 | 1.59E-10 | 4.75E-11 |

**Table S2. $Bi_xO_ySe_z$-WS$_2$ measured diffusion coefficients from Fick's 2$^{nd}$ Law of Diffusion fitting.** Intercalation and deintercalation cycling is shown in Figure S12. The diffusion coefficients are comparable to those measured with $Bi_xO_ySe_z$-WSe$_2$, suggesting the TMD has a minimal impact on the diffusion speed.

## Section S11. AFM scan of MBE grown Bi₂Se₃

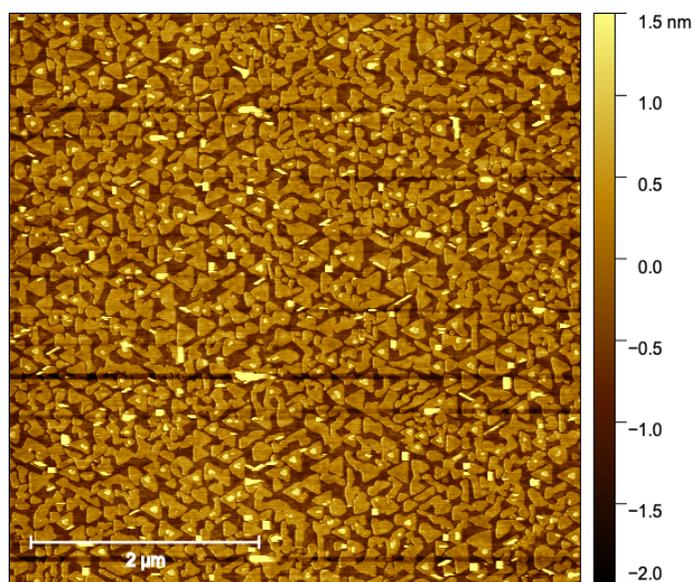

**Figure S14. AFM scan of as-grown MBE Bi₂Se₃.** The samples grew ~5QL tall. The top layer grew incomplete and is composed of triangular islands approximately ~100nm in size. Other defects are observed, which are maximum 6.5nm tall. The morphology observed in AFM is in agreement with STEM imaging, where triangular features were also observed.